\documentclass[prb, article]{revtex4}
\usepackage{epsfig}
\usepackage[centertags]{amsmath}
\usepackage{graphicx}

\newcommand{\one}{\texttt{1}}
\newcommand{\G}{\mathcal{G}}

\newcommand{\F}{\mathcal{F}}

\newcommand{\eps}{\varepsilon}

\newcommand{\w}{\omega}
\newcommand{\up}{+}
\newcommand{\down}{-}
\newcommand{\bra}{\rangle}
\newcommand{\ket}{\langle}
\newcommand{\hwo}{\frac{\tilde{\omega}_0}{2}}
\newcommand{\wo}{{\tilde{\omega}_0}}
\newcommand{\ka}{|a\rangle}
\newcommand{\kb}{|b\rangle}
\newcommand{\ct}{\textrm{cos}~\theta}
\newcommand{\st}{\textrm{sin}~\theta}
\newcommand{\ctt}{\textrm{cos}(2\theta)}
\newcommand{\stt}{\textrm{sin}(2\theta)}
\newcommand{\cttv}{\textrm{cos}^2 (2\theta (v))}
\newcommand{\sttv}{\textrm{sin}^2 (2\theta (v))}
\newcommand{\opt}{\left(| c_a |^2-|c_b | ^2 \right)}
\newcommand{\Rtun}{Re(2 c_a c_b ^*)}
\newcommand{\Ctun}{Im(2 c_a c_b ^*)}
\newcommand{\se}{\textrm{sin}(2\epsilon t)}
\newcommand{\ce}{\textrm{cos}(2\epsilon t)}

\begin{document}

\title{High-order Harmonic Generation and Dynamic Localization in a driven two-level system, a non-perturbative solution using the Floquet-Green formalism}

\author{D.F. Martinez}
\affiliation{Max-Planck-Institut f\"{u}r
 Physik komplexer Systeme, N\"{o}thnitzer Stra\ss e 38, Dresden 01187}


\begin{abstract}
We apply the Floquet-Green operator formalism to the case of a
harmonically-driven two-level system.  We derive exact
expressions for the quasi-energies and the components of the
Floquet eigenstates with the use of continued fractions. We study
the avoided crossings structure of the quasi-energies as a
function of the strength of the driving field and give an interpretation
in terms of resonant multi-photon processes.  From the
Floquet eigenstates we obtain the
time-evolution operator. Using this operator we study Dynamic Localization and High-order
Harmonic Generation in the non-perturbative regime.
\end{abstract}
\maketitle
%
%
\section{Introduction}

The driven two-level system is a particularly important model in
physics, since it has proved to be very useful in describing many aspects of
the interaction of matter with an  electromagnetic field.
In the field of atomic physics, this system was traditionally studied using the
Rotating Wave Approximation (RWA) \cite{Shore}, in
which the counter-rotating term of the harmonic driving is
neglected. This approximation is valid only for amplitudes
of the driving field that are small compared to the energy difference between the
atomic states most affected by the driving field. This was typical for
experiments until the 1960's. Presently, technical improvements in
laser technology allow experimenters to produce
very strong electromagnetic fields (laser pulses) which can produce interesting
non-linear effects in their interaction with matter.

The excitation of a spin 1/2 atom or electron in a magnetic field
interacting with a microwave field is another example where the
strength of the driving field (RF radiation) can not be
assumed to be small.

Also, due to the progress in lithographic techniques and
tunable lasers in the terahertz domain, it has become possible to
construct semiconductor quantum well structures driven by a strong
electromagnetic field.  In these systems, several interesting results have been obtained, including, coherent suppression of tunneling
[also called Dynamic Localization (DL)] \cite{Grossmann}, collapse
of mini-bands in super-lattices \cite{Holthaus92}, absolute negative
conductance \cite{Aguado,DakhnovskiiMetiu}, photon-assisted
tunneling \cite{Aguado96} and AC stark effect \cite{Holthaus94}.

The driven two-level system has also been used to describe an atom moving
through a Fabry-P\'{e}rot cavity.\cite{Meystre} In this system, the strong
coupling regime is already accessible both in the microwave and in the visible regime.

In all of the systems mentioned before, where strong fields are
involved, the "counter rotating" term in the time-dependent part
of the Hamiltonian does contribute significantly. Therefore,
different theoretical approaches beyond RWA are required. One such
method was first introduced by Autler and Townes \cite{Autler}. Using Floquet's theorem and continued fractions, they derived a
general solution in order to investigate the effect of a radio
frequency field on the \textit{l}-type microwave absorption line
of the doublet $J=2\rightarrow 1$ of the molecules of gaseous OCS.

Later, Shirley \cite{Shirley} employed Floquet
theory to reduce the solution of a periodic
time-dependent Hamiltonian to the problem of diagonalization of a
time-independent matrix. This matrix was then used to calculate
eigenvalues and transition probabilities for a driven two-level
system and also to derive higher order (beyond linear RWA)
analytic approximations valid in the case of weak driving. Later
approaches to this problem, usually within the context of a spin
system in a magnetic and a RF field, have involved the use of
continued fractions \cite{Ruyten,Stenholm,Agarwal}, although the
role played by Floquet theory in those approaches (and in
particular the role of the quasi-energy) is not clear.

A different kind of analytical expression for the quasi-energy of
the system, which does not involve continued fractions was
reported by Zhao \cite{Zhao}. It makes use of group theoretical
arguments for its derivation and involves the evaluation of
complicated infinite sums. Apart from its intrinsic theoretical
value, this solution does not provide a practical way to study
this system.

There have been a variety of approaches used to obtain approximate
analytic expressions. Most of them have been
obtained in the high frequency limit \cite{Wang96,Dakhnovskii,Sachetti}, for
$\omega \gg \omega_0$ or in the
strong field limit \cite{Frasca,Barata}, for $v \gg \hbar \omega$. Here
$\omega$ refers to the frequency of the driving,$\omega_0$ is the energy
difference between the two levels of the unperturbed Hamiltonian and $v$ is the strength of the driving.  One of the
most interesting features of a system described by these
perturbative solutions is that at some specific values of the
driving field strength (zeros of the zeroth-order Bessel
function) Dynamic Localization (DL) occurs, which in the context
of quantum wells means that despite a nonzero tunneling
probability between the wells, a system initially prepared in one
of the wells will remain in that well indefinitely. A
manifestation of DL [also called Coherent Destruction of
Tunneling (CDT)] was first found in a tight binding potential
with an applied AC electric field. \cite{Dunlap} DL in quantum
wells was first studied by Gro\ss mann el al. \cite{Grossmann}

In a paper by Ivanov \cite{Ivanov} in which High-order Harmonic Generation
(HHG) by diatomic molecular ions is considered, and then more
recently in works by Santana et al.\cite{Santana} and also Delgado
and Gomez Llorente \cite{Delgado}, analytical expressions in terms of Bessel
functions were found, valid in the
(perturbative) regime where DL occurs. This regime can be
characterized by the condition $\omega_0 / \omega \ll 0$ for
any $v$, or $\omega_0 /\omega \ll \sqrt{v/\hbar\omega} \gg 1$. We will
say that whenever any of these two conditions is satisfied, the
system is in the Dynamic Localization Regime (DLR). When that regime is
reached, our results converge to the known Bessel-function expressions.

In this work we present a fully developed Floquet-Green formalism
for time periodic systems, which we then use to derive an exact
solution for the two level system with harmonic driving. This
exact solution is expressed in terms of continued fractions,
suitable to the study of the regime where neither the RWA nor any
other perturbative approach works. Our aim is to present a whole
picture of this system, of its eigenvalues and eigenstates in the
different regions of its parameter space and to obtain from the general
solution a better understanding of the two main features of
this system, namely, Dynamic Localization (DL) and High-0rder Harmonic
Generation (HHG).

In section II we derive the Floquet-Green operator formalism and
establish a connection between the poles of
this operator and the components of the Floquet eigenstates.

In section III we apply this formalism to the specific case of the
two-state system with harmonic driving.  By making use of
continued fractions, we obtain an expression for the quasi-energies. We study
 the structure of avoided-crossings in the quasi-energies and interpret them 
in terms of multi-photon processes. We also derive here the Floquet eigenstates (FES) of the
system and study their dependence on the amplitude of the driving field for
different values of the two-level energy difference.
The time-evolution operator is constructed using the FES.

In section IV, using the above mentioned operator, we study Dynamic
Localization (DL), and in section V we study High-order Harmonic Generation (HHG).

In Section VI we summarize our findings and give some concluding remarks.

\section{Floquet-Green operator for harmonic driving}\label{FGc1}

The pioneering work of Shirley \cite{Shirley} and Sambe \cite{Sambe}
laid down the theoretical foundations for a complete treatment of
time-periodic potentials, based on the same mathematical tools
already developed for time-independent potentials. Of great
importance among these tools is the Green's function, whose
definition and application for time-periodic systems has not been
clear until recently.  A Floquet-Green function method for the
solution of radiative electron scattering in a strong laser field
was introduced by Faisal\cite{Faisal}.  In this work we
present a fully developed picture of the Floquet-Green's operator
formalism for general time-harmonic Hamiltonians that was
introduced by the author in a previous work \cite{Martinez03}, and apply it to the case of a two-state system with harmonic driving.

We start by considering a Hamiltonian of the general form:

\begin{equation}
H(t)=H_0 + 2 V \textrm{cos}(\w t),
\label{Hamiltonian1}
\end{equation}
were $H_0$ and $V$ are Hermitian operators in the Hilbert space ($\mathcal{H}$) of
the system. 
Because of the periodicity of the
Hamiltonian, according to Floquet's theorem, the solutions to
Schrodinger's equation $i\hbar\frac{\partial}{\partial t}|\Psi
(t)\bra=H(x,t)|\Psi(t)\bra$ are of the form
\begin{equation}
|\Psi^e(t)\bra=e^{-i e t/\hbar} |\phi^e(t)\bra,
\label{Psiphi}
\end{equation}
where $|\phi^e(t)\bra=|\phi^e(t+\frac{2\pi}{\omega})\bra$.

Inserting this into Schrodinger's equation one arrives at the
eigenvalue equation
\begin{equation}
H^{F}(t)|\phi^e(t)\bra= e |\phi^e(t)\bra ,
\label{Feigenvaleq1}
\end{equation}
where $H^{F}(t)$ is defined as
\begin{equation}
H^{F}(t)\equiv H(t)-i\hbar\frac{\partial}{\partial
t}.\label{floquetHam1}
\end{equation}

As pointed out by Sambe \cite{Sambe}, since Eq.(\ref{Feigenvaleq1})
is an eigenvalue equation, it can be solved using the standard
techniques developed for time-independent Hamiltonians, provided
we extend the Hilbert space to include the space of time-periodic
functions. In this extended space, the time parameter can be
treated as another degree of freedom of the system. A similar
concept is used in classical mechanics, and gives rise to the
concept of a "half" degree of freedom when dealing with time
dependent Hamiltonians.

A suitable basis for this extended Hilbert space ($\mathcal{R}$)
is $\{ |\alpha\bra\otimes |n\bra,...\}$, where
$\{|\alpha\bra,..\}$ is a basis for the Hilbert space
$\mathcal{H}$ of the system, and we define $\ket t|n\bra=e^{-in\w
t}$, with $n$ integer. Clearly $\{|n\bra ,..\}$ spans the vector
space ($\mathcal{T}$ of periodic functions,
and therefore, $\mathcal{R=H\bigotimes T}$. In this basis,
Eq.(\ref{Feigenvaleq1}) becomes a matrix eigenvalue equation of
infinite dimension with an infinite
number of eigenvalues. It is not difficult to proof that if $e_i$
is an eigenvalue, with corresponding eigenvector $|\phi
^{e_i}(t)\bra$, then $e_i +m$ is also an eigenvalue (all
quantities are assumed to be in units of $\hbar\w$), with
corresponding eigenvector $|\phi^{e_i +m}(t)\bra$= $e^{im\w
t}|\phi^{e_i}(t)\bra$. Accordingly, the eigenstate corresponding
to eigenvalue $e_i + m$ has the same structure than the eigenstate
corresponding to $e_i$, except that it is
displaced by $m$ on the energy axis. 
Because of this, to find all
the eigenvectors and eigenvalues of the Floquet Hamiltonian one
needs only to consider $-\frac{1}{2}\leq e <\frac{1}{2}$. We will
use the letter $\eps$ to refer to the Floquet eigenvalues
restricted to this interval and call them "quasi-energies". Clearly, any
Floquet eigenvalue $e_i $ can be written as $e_i=\eps_i +p$ for
some $-\frac{1}{2}\leq\eps_i <\frac{1}{2}$ and some integer $p$.
It can be shown that, in general, there are N distinct quasi-energies (except
for accidental degeneracies) if the
Hilbert space $\mathcal{H}$ is N dimensional.

This periodic structure in the eigenvalues does not mean that the
"replica" eigenstates have no relevance; they are also valid
solutions of Eq. (\ref{Feigenvaleq1}) and are essential for
completeness in the extended Hilbert space
$\mathcal{R}$ \cite{Dresse99}. They also allow us to understand some
features of the quasi-energies of the system in terms of avoided
crossings between "replica" eigenstates.

The Floquet-Green operator
for the Floquet Hamiltonian in Eq. (\ref{Feigenvaleq1}), is
defined by the equation \cite{Martinez03},

\begin{equation}
[\one
E-H^{F}(t')]\G(E,t',t")=\one\delta_{\tau}(t'-t'')~~,\label{definitionG}
\end{equation}
where $\delta_{\tau}(t)$ is the $\tau$-periodic delta function
($\tau=\frac{2\pi}{\w}$).

In terms of the complete
(infinite) set $\left\{ |\phi^{e_i}(t)\bra\right\}$  of eigenfunctions of the
Floquet-Hamiltonian (Eq.\ref{floquetHam1}), the solution for
Eq.(\ref{definitionG}) is

\begin{equation}
\G(E,t',t'')=\sum_i \frac{|\phi^{e_i
}(t')\bra \ket \phi^{e_i}(t'')|}{E-e_i}~.
\end{equation}

From the previous discussion about the eigenvalues and
eigenfunctions of the Floquet Hamiltonian, we can write the
Floquet-Green operator entirely in terms of the eigenfunctions
corresponding to values {$e_i$} between $-\frac{1}{2}$ and
$\frac{1}{2}$. We will denote these eigenvalues as $\eps_\gamma$.
Using this, the Floquet-Green operator can be written as
\begin{equation}
\G(E,t',t'')=\sum_\gamma \sum_p e^{ip\w
(t'-t'')}\frac{|\phi^{\eps_\gamma }(t')\bra\ket
\phi^{\eps_\gamma}(t'')|}{E-\eps_\gamma-p}~~,
\end{equation}
where $\gamma=1,...N$ for $\mathcal{H}$ being N-dimensional, and
$p=-\infty,...,\infty$.

Operating on both sides of this equation with
$\frac{1}{\tau^2}\int_0 ^\tau \int_0 ^\tau e^{im\w t'}e^{-in\w
t''}dt''dt'$ we obtain
\begin{equation}
\G_{m,n} (E)=\sum_{\gamma,p}
\frac{1}{E-\eps_{\gamma}-p}|\phi^{\eps_{\gamma}}_{m+p}\bra\ket\phi^{\eps_\gamma}_{n+p}|,
\label{Gandeigenveccomp1}
\end{equation}
where
\begin{equation}
\G_{m,n}(E)=\frac{1}{\tau^2}\int_0 ^\tau \int_0 ^\tau e^{im\w
t'}e^{-in\w t''}\G(E,t',t'')dt''dt'~, $$and$$
|\phi^{\eps_{\gamma}}_{m}\bra=\frac{1}{\tau}\int_0 ^\tau e^{im\w
t'}|\phi^{\eps}(t')\bra dt'.
\end{equation}

For $m=n=0$,
\begin{equation}
\G_{0,0}(E)=\sum_{\gamma,p}
\frac{1}{E-\eps_{\gamma}-p}|\phi^{\eps_{\gamma}}_{p}\bra\ket\phi^{\eps_\gamma}_{p}|~.
\label{G00}
\end{equation}

This last equation shows us the relationship between the residue of the operator $\G_{0,0}(E)$ at the pole
$\eps_{\gamma}+p$ and the Fourier component
$|\phi^{\eps_{\gamma}}_{p}\bra$ of the Floquet eigenstate
$|\phi^{\eps_{\gamma}}(t)\bra$. Notice that
$|\phi^{\eps_{\gamma}}(t)\bra=\sum_p e^{-ip\w
t}|\phi^{\eps_{\gamma}}_{p}\bra$. From this we conclude that,
\textit{corresponding to each  pole at $E=\eps_\gamma +p$ of the
Floquet-Green operator $\G_{0,0}(E)$ there is an oscillating term
of the form $e^{-ip\w t}$ in the Floquet eigenstate specified by the quasi-energy $\eps_\gamma$}.

Let us go back to the task of finding an explicit solution to
the Floquet-Green operator in Eq.(\ref{definitionG}). Applying
$\frac{1}{\tau}\int_0 ^\tau \int_0 ^\tau e^{im\w t'}e^{-in\w
t"}dt"dt'$ on both sides of this equation, and using
Eq.(\ref{floquetHam1}) together with the definitions above we get

\begin{equation}
[\one (E+m) - H_0]\G_{m,n} - V (\G_{m+1,n} + \G_{m-1,n})=\one
\delta_{m,n}
\label{Gmnequation}
\end{equation}

The explicit solution of this equation, in terms of matrix
continued fractions follows from \cite{Martinez03}. 
The resulting expression for the operator $\G_{0,0}(E)$ is

\begin{equation}
\G_{0,0}(E)=(\one E - H_0 -V_{\rm{eff}}(E))^{-1},
\label{almostfinalresult}
\end{equation}
where
\begin{equation}
V_{\rm{eff}}(E)= V_{\rm{eff}} ^{+}(E ) + V_{\rm{eff}}^{-}(E )~~,
$$with$$
V_{\rm{eff}} ^{\pm}(E)= V \frac{1}{\displaystyle E \pm 1-H_0 -V
\frac{1}{\displaystyle E\pm 2-H_0 -V\frac{1}{~~\vdots~~}V}V}V~~,
\label{Veff}
\end{equation}

We now specialize this result for the case of a two state system
driven by a classical single frequency potential.
\section{Harmonically driven two-level system}\label{FGc2}

In this section we apply the formalism previously presented to
the case of a driven two-level system of the
form
\begin{equation}
H=\hwo \sigma_{z} + 2v\textrm{cos}(\omega t)\sigma_x ~,
\label{2Sham}
\end{equation}
where $H$, $\wo$ and $v$ are given in units of $\hbar\omega$. In the context of an electron interacting with an electric field, the
amplitude $2v$ corresponds to the (static) dipole moment of the electron times
the applied electric field, i.e. $2v= \mu E_0 / \hbar\omega$, (the factor of
two has been introduced for convenience) and the classical electric field is of the
form $E(t)=E_0 \textrm{cos}(\omega t)$. In a double quantum-dot realization
of  this Hamiltonian, the non-driven part, $ \hwo \sigma_z$,
describes the tunneling between the
states localized on each dot, $\{|1\bra , |2\bra \}$. This tunneling gives rise to a
splitting of
$\hbar \omega_0 = \wo \hbar \omega$ between the energy levels $-\frac{\hbar
  \omega_0}{2}
 , +\frac{\hbar \omega_0}{2}$, whose corresponding
eigenstates we denote by $\{\ka, \kb \}$, and where  $\kb =\frac{1}{\sqrt{2}}(|1\bra
  + |2\bra )$, and   $\ka =\frac{1}{\sqrt{2}}(|1\bra
  - |2\bra )$. The Hamiltonian in Eq.(\ref{2Sham}) is written in the basis 
$\{\ka,\kb\}$.
For an atomic system, $\hwo \sigma_z$ describes two  bound states of the
atomic (or molecular) potential, or, in the context of ionizing systems it
could also be thought of as describing a bound state and
the continuum \cite{Kaplan,Dimitrovski}. An electron moving in a
Fabry-P\'{e}rot cavity \cite{Meystre} or a 1/2 spin system with an applied magnetic  and
microwave field \cite{Ruyten,Stenholm,Agarwal}, are also examples of systems where this model has been used. 

Continued fractions methods have been used before to study this
system, specially in the context of spin 1/2 systems in a magnetic
and RF field, starting with the seminal work of Autler and
Townes\cite{Autler}. Also, a previous treatment of this problem
where the time dependent Green's function and continued fractions
were used, although without reference to the Floquet formalism can
be found in the work by Gush and Gush \cite{Gush}. There have been several
perturbative approaches to this problem, as it was mentioned before. Here, we will focus on the properties of the
exact solution, and in particular we will study its dependence on
the strength of the driving field and the connection with HHG and DL.

To find the exact solution to this Hamiltonian we need to evaluate
Eqs. (\ref{almostfinalresult},\ref{Veff}), for the case
$V=v\sigma_x $ and $H_0 =\frac{\omega_0}{2} \sigma_z$. It can be
checked easily that, given the off-diagonal form of $V$ and the
diagonal form of $H_0$, the matrices $V_{\rm{eff}} ^{\pm}(E)$ are diagonal.
Its diagonal components will be denoted respectively by $\{v_a
^{\pm}(E),v_b ^{\pm}(E)\}$. The dynamic effective-potential
$V_{\rm{eff}} (E)$ is therefore also diagonal and of the form
\begin{equation}
V_{\rm{eff}}(E)=\left[ \begin{array}{rr}
                 v_a ^{+}(E)+v_a ^{-}(E)& 0 ~~~~~~~~  \\
               0~~~~~~ &  v_b ^{+}(E)+v_b ^{-}(E)
                        \end{array}\right].
\label{DEP1-2states}
\end{equation}
The  $v_{a/b}^{\pm}$ functions are given by the following coupled
recursive relations
\begin{equation}
v_a ^{\pm}(E)= \frac{v ^2}{ E\pm 1-\hwo-v_b ^{\pm}(E\pm 1)},~~~
v_b ^{\pm}(E)= \frac{v ^2}{ E\pm 1+\hwo-v_a ^{\pm}(E\pm 1)}.
\label{vavb}
\end{equation}
Replacing $v_{a}^{\pm}$ into $v_{b}^{\pm}$ (and vice versa) we
obtain the de-coupled recursive relations
\begin{equation}
v_a ^{\pm}(E)= \frac{v ^2}{ \displaystyle E\pm 1-\hwo
-\frac{v^2}{\displaystyle E\pm 2+\hwo-v_a^\up (E\pm 2)}} ~~, ~~~~~~~~~
v_b ^{\pm}(E)= \frac{v ^2}{ \displaystyle E\pm 1+\hwo
-\frac{v^2}{\displaystyle E\pm 2-\hwo-v_b^\up (E\pm 2)}}.
\label{v'sdecoup2state}
\end{equation}
From Eq.(\ref{almostfinalresult}), the solution for $\G_{0,0}(E)$
is also a diagonal matrix of the form
\begin{equation}
\G_{0,0} (E) =\left[ \begin{array}{rr}
                 g^a (E) & 0   \\
               0 & g^b (E)
            \end{array}\right],
            \label{greens2state}
\end{equation}
with
\begin{equation}
g^a (E)=\frac{1}{\displaystyle E+\hwo-v_a^\up (E)-v_a^\down (E)}~,
$$ $$ g^b (E)= \frac{1}{\displaystyle E-\hwo-v_b^\up
(E)-v_b^\down (E)}~.
\label{gagb}
\end{equation}

It can easily be seen from Eq.(\ref{v'sdecoup2state}) that this
system possesses the symmetry $v^{\pm} _a (-E) = -v^{\mp} _b (E)$,
which from Eq.(\ref{gagb}) gives us $g_a (-E)= - g_b (E)$. This
means that to find the two quasi-energies of the system one only
has to look for a pole in either $g_a (E)$ or $g_b (E)$ in the
range $-0.5 \leq E \leq 0.5$; the other pole is symmetrically
located on the opposite side of the (quasi) energy axis.

\subsection{Poles of $\G_{0,0}(E)$ and the quasi-energies of the system}

As discussed before, the quasi-energies of the system can be
obtained from the poles in $g_a(E)$ and $g_b(E)$. We showed that
it is sufficient to study the poles in only one of these functions
since for any pole in $g_a(E)$ at $E=E^*$ there is a corresponding pole in
$g_b(E)$ at $E=-E^*$. This implies the existence of two
quasi-energies in the system, $\eps, -\eps$. The quasi-energy
$\eps$ can be obtained from the solution to $g_b(E)^{-1} = 0$ in
the interval $-0.5<E<0.5$. From
Eqs. (\ref{v'sdecoup2state}),(\ref{gagb}), the poles of $g_b(E)$
satisfy the equation,

\begin{equation}
\delta= \frac{v ^2}{ \displaystyle \\wo
-\frac{v^2}{\displaystyle \delta+2-\frac{v^2}{\displaystyle
\delta+3+\wo-\frac{v^2}{\displaystyle
\delta+4-\frac{v^2}{\vdots}}}}}+    \frac{v ^2}{ \displaystyle
\delta-1+\wo -\frac{v^2}{\displaystyle
\delta-2-\frac{v^2}{\displaystyle
\delta-3+\wo-\frac{v^2}{\displaystyle
\delta-4-\frac{v^2}{\vdots}}}}}, \label{mastereq}
\end{equation}
where $\delta=E-\hwo$. The quasi-energy is obtained as $\eps =
(\hwo+\delta) ~mod ~1$.

In Fig. 1 we show one of the
quasi-energies of the system, for different values of $\wo$. These functions are plotted as a function of the driving
strength $v$. We have included the
resonant case $\wo =1$ and several values in the high frequency
regime where $\wo<1$. In the limit when $\wo << max(1,\sqrt{v})$, the
function $\eps (v)\approx \hwo J_0 (4v)$. This result was first
obtained by Shirley \cite{Shirley}. It is precisely in this
regime where it was shown \cite{Santana,Delgado} that DL appears
as a dominant feature. As $\wo$ decreases we observe a very quick
convergence to the approximate result  $\eps (v)\approx \hwo J_0
(4v)$. Already for $\wo=0.25$, the difference between the exact
result for $\eps(v)$ and the approximation cannot be observed at the scale of the graph. For $\wo=0.1$
we plotted both $\eps(v)$ and $-\eps(v)$ to make the
location of the zeros of $J_0 (4v)$ more visible. As we can also see in Fig.2,
increasing $\wo$ always displaces the zeros of $\eps(v)$ towards
smaller values. This behavior was also observed by Villas-B\^{o}as
et al.\cite{villas} in the case of a double quantum well with an
intense AC electric field and an applied magnetic field.
Notice also that, as it is easy to derive analytically, the only
case when the initial slope of the quasi-energy is not zero is for
the resonant case. \cite{approx}

In Fig. 2 we show the quasi-energy as a function of the driving
strength $v$, for low frequencies, where $\wo \geq 1$. The
resonant case is included for comparison. Also, the dotted line
shows the functions $\pm 0.05 J_0 (4V)$  in order to indicate the location
of the zeros of this Bessel function.  As one increases $\wo$, the amplitude and sharpness of the oscillations
around $\varepsilon = 0$ increases, the location of the zeros
shifts to the left, and the parabola that describes (except in the
resonant case) the initial behavior of the curve opens up. For
$\wo \gtrsim 2$  one can get from Eq.(\ref{mastereq}) that the
initial behavior of the quasi-energy is parabolic and has the form
\begin{equation}
\delta(v) \approx \frac{2\wo }{\wo^2-1}v^2
\end{equation}
This remains a good approximation for all values of $v$ until
$\delta(v) \approx \frac{\wo-1}{4}$, and provided $\epsilon (v) $
does not reach the edges of the Brillouin zone (-0.5,0.5) before
that. If it does so, then, the quasi-energy experiences an
avoided crossing, whose sharpness is proportional to $\wo$ and
decreases with the value of $v$ at which the anti-crossing occurs.

\subsection{Avoided crossings}

In Fig.3 we can see that the structure of the quasi-energy is
determined by a regular sequence of avoided crossings which get
broader as the strength of the driving field $v$ is increased (resonant case shown
in the figure).

Let us consider the first
avoided crossings, the ones along the line marked "3 ph. res" in Fig. 3.
To zero order in $v$, the eigenvalues and eigenvectors of the
Floquet-Hamiltonian are the state vectors
${|a\rangle|n\rangle,|b\rangle|n\rangle}$, for any integer $n$. In this basis,
the time-dependent potential can be written as
$V=v\sigma_x (a+a^\dagger )$.
When the effect of one photon processes (in the RWA) is included,
one obtains that the new set of eigenstates is $\{|n,\pm
\rangle \}$, for any integer $n$, and where $|n,\pm
\rangle =
\frac{1}{\sqrt{2}}(|a\rangle|n\rangle\pm|b\rangle|n-1\rangle)$.
It's corresponding eigenvalues are ${e}_n ^\pm = \mp \frac{1}{2} +n\pm v$.
This one photon process is in a sense an avoided crossing, with
a width that goes to zero as $v\rightarrow
0$. One can see that, near $v\approx 1$, the eigenvalues
corresponding to the eigenstates $|n, \pm\rangle$ and $|n \pm
2, \mp \rangle$ would cross if only fist order processes where
allowed. To second order this degeneracy is not lifted since
$\langle n \pm 2, \mp| V^2 |n,\pm \rangle =0$. Only when third order
processes are included, this degeneracy is lifted, since
it is easy to check that $\langle n \pm 2,\mp | V^3 |n
\pm \rangle \neq 0$. Third order processes are therefore involved in the avoided crossings near $v\approx 1$.

For the resonant case, after the first  avoided crossing ($v\approx 1$), for example between the states $|{1},-
\rangle$ and $|{-1},+ \rangle$, and because of considerable overlap between
the 3 photon resonance and the 5 photon resonance, this states do not completely exchange their
identities, as it would be the case in a well isolated avoided
crossing.  An "identity exchange" between Floquet states can clearly be observed for values of $\wo$ and $v$
for which there is a sharp avoided crossing (i.e. $\wo =~3.0$ in
Fig.5). In Fig.3, for the resonant case we can see that even though the structure of the eigenstates
after the first avoided crossing is not necessarily of the form
$|n\rangle\rangle$, the points where the quasi-energies of these "first-order" FES cross, come close to the actual location of
the avoided crossings. This is an indication that those states are
dominant in the structure of the eigenstates, a fact that can be checked by
comparing the actual structure of the FES in Fig. 6, at some amplitude
$v$, with the corresponding "first order" FES eigenvalues that intersect at the closest
avoided crossing in Fig. 3. For
$v\approx 5$  the peak components in the FES are only one unit of energy
further out that the expected value from the simple picture in Fig.3. It is
an interesting feature of this system the fact that the FES obtained for small
driving amplitudes constitute the "bare bones"
structure of the FES and to a good extent they determine the avoided
crossings structure of the quasi-energies (at least for a larger-than-expected range of driving
field amplitudes).

This avoided crossings structure, to a certain approximation, can
be thought of as mostly due to odd-photon processes between FES. Even more, we will
argue that at least for a range of $v$ ($0< v \lesssim 4$ for the resonant case), the
main contribution to the avoided crossings is a three photon process for
the first avoided crossing, 5 photon for the second avoided crossing, and so on (in the resonant case). For any  value
of $\wo$ we find that the number of photons involved in the first
resonance depends on the initial distance between the two unperturbed levels ($\wo$). If
for example $\wo =3$ the first avoided crossing will be the 5 photon
resonance (Fig.5); for $\wo =5$  it will be the 7 photon resonance, etc. In general we can write that, for $\wo\geq 1$ the order of the
first resonance is $2~ \rm{Int}( \frac{ \wo -1}{2} )+3$. The avoided crossings
picture holds for any $\wo$ provided $v$ is not too big; an estimate is
$v\lesssim 4 \wo ^2$. Above this, the overlap between different photon processes is too big for our simple picture to apply. 

In Fig.3. it can be seen that the states that intervene at each avoided
crossing are of the form 
$|2n+1,-\bra$ and $|2n+1-2m,+ \bra$, for $m=1,2,3...$. It is easy to check that $$\ket
2n+1,-| V^q |2n+1-2m,+ \bra \neq 0, ~~ \textrm{ only for}~~ q=2m-1, 2m+1,
...$$
This is a good indication that these avoided crossings can be interpreted as multi-photon
resonances that involve odd numbers of photons. From Fig. 3, we see that at
the avoided crossing A, from the interacting states $|1,-\bra$ and
$|-1,+\bra$, there should be a three photon exchange between $|a\bra|1\bra$ and
$|b\bra|-2\bra$ and a 1 photon exchange between  $|a\bra|-1\bra$ and
$|b\bra|0\bra$. From the exact result for the components of one of the FES
shown in Fig. 6 we
see that the dominant component before the crossing is $|a\bra |1\bra$ and
therefore, at $v=1$ (which is near the crossing), the two dominant components should be $|a\bra
|1\bra$ and $|b\bra |-2\bra$; this corresponds to a three photon
process. Near the avoided crossing B in Fig.3, where $v \sim 2$, we can see from Fig.5 that the dominant components are $| a,3\bra$ and $|
b,-2\bra$ which corresponds to a 5 photon exchange. For values of $\wo$ and
$v$ for which the avoided crossings are sharper one gets an even  clearer
picture of those avoided crossing as resulting from the exchange of a specific
odd number of photons between Floquet states.
 
It is interesting to see how this description of avoided crossings
and photon processes is reflected by the structure of the
continued fractions in Eq.(\ref{mastereq}) and how the different
terms in it contribute to the features of the quasi-energy as a
function of the driving amplitude. From examination of
Eq.(\ref{mastereq}) it is not difficult to see that
for small $v$, each term of the form $\delta+\alpha$ gives a
solution near $\delta=-\alpha$. To find accurately the solution of
$\delta$ for a particular value of $v$ one does not need to (nor
can) include an infinite number of terms in the C.F. An important
question is therefore, how many terms to include and why? To study
this, we first propose a way to truncate Eq.(\ref{mastereq}) so that
 we obtain, in each case, a symmetrical set of solutions. For
illustration we will continue focusing on the particular case of resonance
($\wo = 1$), as it is straight forward to apply the same ideas shown
here for  other values of $\wo$.

Eq.(\ref{mastereq}) can be put in the form
\begin{equation}
F^+ (E) F^- (E) = v^2, \label{symmastereq}
\end{equation}
with
\begin{equation}
F^+ (E) = \delta-\frac{v ^2}{ \displaystyle \delta + 1+\wo
-\frac{v^2}{\displaystyle \delta + 2-\frac{v^2}{\displaystyle
\delta +3+\wo -\frac{v^2}{\vdots}}}}~,~~~F^- (E) =
\delta-1+\wo-\frac{v ^2}{ \displaystyle \delta - 2
-\frac{v^2}{\displaystyle \delta - 3+\wo-\frac{v^2}{\displaystyle
\delta -4 -\frac{v^2}{\vdots}}}}.
\end{equation}
To understand the effect of the different terms in the C.F. we
will truncate the functions $F^\pm$ at different levels and plot
the solutions for $\delta$. $F^\pm _n$ refers to the truncated
version of $F^\pm$ where $v^2$ appears $n$
times\cite{convergence}, i.e.  $$F^+ _0= \delta, ~~~F^+
_1= \delta-\frac{\displaystyle v^2}{\displaystyle \delta+ 1+\wo},
~~~~F^+ _2= \delta-\frac{\displaystyle v^2}{\displaystyle \delta
+1+\wo-\frac{\displaystyle v^2}{\displaystyle \delta + 2}},~~~ ....~, $$
and likewise for $F^- _n$.

In the different panels in Fig. 4 we plot the solutions for
$\delta(v)$ as obtained from the equation $F^+ _n F^- _n =v^2$ for
$n=0,1,2,3$ [ panels a) through d)],  and only for the resonant
case $\wo=1$. We also plot with a dashed line and for comparison,
the exact result for $\delta$ (only the two smallest roots are
shown). Also, in front of each plot there is an insert with a
diagram with lines and arrows. The lines represent each term that
has been included in the continued fractions $F^+ _n$ and $F^-
_n$. As we mentioned before, each term of the form $\delta+\alpha$
contributes with an eigenvalue curve that originates at
$\delta=-\alpha$. There are two kinds of terms of that form in the
C.F.'s of Eq.(\ref{mastereq}): $\delta\pm 2n$ and $\delta +\wo \pm
(2m+1)$. The first ones contribute, at $v=0$, with a solution at
$\delta=\mp 2n$ and therefore a pole at $E= E_b\mp 2n$. The second
kind gives a solution for $\delta=-\wo \mp (2m+1)$ which
contributes with a pole at $E= E_b -\wo \mp (2m+1) = E_a \mp
(2m+1)$.  To the lines on the left we associate the
states  $|a,2n+1\bra$ . We will label this
lines as $a + 2n+1$, for some n; the lines on the right we
associate with the particle-field states $|b,2n\bra$ and
label them as $b + 2n$. A similar picture would result from
studying the poles of $g_a (E)$. 

When two such terms in the CF are included, then we see, in the corresponding panel in Fig.4,  a pair of solutions, which at $v=0$ converge to $\delta=2n$ and
split for $v>0$. We know, from previous discussions, that the upper one should
correspond to the FES, $|2n+1,+\bra= |a,2n+1\bra + |b,2n\bra$, and the lower one
to $|2n+1,-\bra= |a,2n+1\bra - |b,2n\bra$. The arrows at the end of the
lines show the direction in which the corresponding eigenvalue would move when $v$ is
increased.  The repulsion between the eigenstates $|2n+1,\pm\bra$ is due to the exchange of one photon
between $\ka$ and $\kb$. This is the main process at work in
panels a) and b). In panel c) two new processes enter the picture:
As the levels move apart due to the one photon resonance, another
resonance (A) can now occur when the levels  $|3,-\bra$ and $|1,+\bra$ mostly
due to the transfer of three photons between $|a,3\bra$ and $|b,0\bra$.
This process appears for the first time for $F^\pm _2$ when the term $|a,3\bra$ is included
in the CF. A replica of this three
photon resonance occurs also between levels $|1,-\bra$ and
$|-1,+\bra$. 

When $|a,3\bra$ and $|b,-2\bra$ is included, the second process
that becomes available is the
five photon resonance between levels $|3,-\bra$ and $|-1,+\bra$ (after this
level have interacted with $|1,+\bra$
 and $|1,-\bra$ respectively). This resonance becomes more clearly
defined in panel d) where more levels are included. Even though,
as compared to c), no new resonances appear in d), the inclusion
of levels $|a,-3\bra$ and $|b,4\bra$ contributes to sharpen the
resonances already visible in c) and in particular the 5 photon resonance B. This shows that, as mentioned before,
there are several odd-photon processes that contribute to a particular avoided
crossing. There is however, an odd number of photons that seems to contribute
the most and without which, if the corresponding levels are not included, the
corresponding avoided crossing does not seem to occur. That specific number of photons
is shown on the top of Fig.3. For $v > 4$ this pictures begins to
falter, and the avoided crossings begin to depend more on a higher
number of photons than what would be predicted from Fig.3.  For $v=5$ for
example, a better description of the crossing would be for a 15 photon
resonant process, not the expected 13. 
For even higher values of $\wo$ this picture holds on a bigger range of the
driving field amplitude, namely, for $v\lesssim 4 \wo^2 $. As mentioned before, one
can see that in Fig.5, for $\wo = 3$,  the first avoided crossing is clearly a 5 photon
resonance.

As an empirical rule, which agrees with the above observations, we found that to obtain an accurate result for the
quasi-energy up to a value $v$ of the driving potential one needs to use the
function $F^{n}(E)$, with $n > \textrm{Int}(2v)$.


\subsection{Components of the Floquet EigenStates}

In the previous subsection we showed how to obtain the diagonal
part of the Floquet-Green operator and discussed its pole structure and
their avoided crossings.  In this part we study the Floquet eigenstates and the general
behavior of them as the amplitude of the driving field changes.
In Appendix A we show that the FES of this system can be written in the form

\begin{equation}
|\phi^{-\eps}(t)\bra=K(t)|a\bra~+ ~Z(t)|b\bra
$$and$$
|\phi^{\eps}(t)\bra= -Z(t)^* |a\bra ~+~ K(t)^* |b\bra ,
 \label{eigenvecfinal}
\end{equation}
with
\begin{equation}
\begin{aligned}
K(t) &=N[\sum_{n=-\infty}^\infty e^{-2in\w t}K_{2n} (-\eps)]~,
\\
Z(t)&=N[\sum_{n=-\infty}^\infty e^{-i(2n-1)\w t}Z_{2n-1}(-\eps) ].
\end{aligned}
\end{equation}
$N$ is a normalization constant and $|K(t)|^2 + |Z(t)|^2=1$. In appendix
A we also
show the explicit form of the components $K_{2n}$, $Z_{2n+1}$, in Eq.(\ref{components}).
It can there be seen that all components $K_{2n}$ and $Z_{2n-1}$ are real, and
therefore $K(0)$ 
and $Z(0)$ are also real numbers. 

In Fig.6 we show the components of one of the Floquet
 eigenstates of the system, $|\phi^{\eps}(t)\bra$, for the resonant case, and for eight different values of the driving field amplitude,
In the upper figure of each panel we show the quantities $Z_{n}$,$K_{n}$ (with
 n odd for $Z_{n}$ and n even for $K_{n}$), and in the lower panels we show the quantities ln$|Z_n
 |$, ln$|K_n| $. 

A noticeable qualitative difference between the behavior of
$Z_{2n-1}$ and $K_{2n}$ is the fact that for stronger values of
$v$, the former becomes more antisymmetric while the last one
becomes more symmetrical around $n=0$. We will comment shortly about
the consequence that this has for the dynamic localization
of the system. The fact that the eigenstates of the system have components
that are harmonics of the driving field frequency $\omega$, translates, as we
will study in the next section, into a (coherent) emission spectrum of
the system which
contains also many frequencies that are multiples of the frequency of the driving
field.  The possibility of exciting a system at a
particular frequency with a strong field, and  producing radiation that includes
many harmonics of the driving frequency (in some cases even
 hundreds of them) is called High Harmonic Generation (HHG) and has
 been studied in many experiments.\cite{eden}
Driven two-level systems have been studied in the past to understand the
general features of HHG such as the existence of a plateau in the spectrum
followed by a sharp cut-off where the amplitudes of the harmonics decrease 
rapidly.\cite{Sundaram,Ivanov,Dakhnovskii,Gauthey97,Kaplan,Gauthey95,Pons}

As one can see directly from Fig. 6 (resonant case), in most of the cases shown, the Floquet eigenstates show a
 particular structure with two clearly distinguishable regions. The first
 region which, in a non-rigorous way we call the "chaotic" one,
can be described as corresponding to the Floquet components
with  $|n| \lesssim N_{e}= 2v-1$ (for $v\geq 1$ in the resonant case). The amplitude of these Floquet components depends strongly on $n$ and on
 $v$, and there are
 frequent sign changes. This gives rise to the
 "plateau" region typical in HHG. The 
second region, or the "regular" one corresponds to the components with
 $|n| > N_{e}$, and it is characterized by having amplitudes that decay exponentially
with $n$. Also, notice that the components $K_n$ and $Z_n$ do not change sign
in that region, with $K_{2n} >0$, and $\textrm{sg}( Z_{2n-1}) = \textrm{sg}(n)$  in the
" regular" region ($\textrm{sg}(x)\equiv x/|x|$).  
For other values of $\wo$, a plateau region was always found whenever $v
 \gtrsim \sqrt{\wo}$ (for $\wo \geq 1$), in
 agreement with a similar expression obtained by Kaplan\cite{Kaplan} in the
 limit $\wo>>1$. For  $\wo<1$, the plateau forms for any $v \gtrsim 1$.

The expression $N_{e} = 2v-1$ can be derived in a simple way.
 First, we look at the conditions to obtain, for $n>0$ a perfectly flat, triple plateau in the components
 of the FES. This means that we will assume $K_{n+2} =Z_{n+1} = K_{n}$ (which
 gives three consecutive components with the same amplitude), for
 some $n$ even. This condition can be achieved if, from Eq.(\ref{components}),
 $v_a^+ (\eps +n) = v_b^+ (\eps +n+1) = v$. From Eq.(\ref{vavb}) we get the
 condition,
$$n=2v-1+\hwo-\eps.$$
If instead, we had chosen the triple plateau to be of the form $Z_{n+2} =K_{n+1} =
Z_{n}$ ( for $n$ odd), then, the equation for $n$ would be
$$n=2v-1-\hwo-\eps.$$
If we now apply the same reasoning for a triple cusp to
occur in the negative energy components, beginning at the component $-n$,  we
get the conditions (for $n$ odd and even)
$$n=2v-1+\hwo+\eps,$$ $$n=2v-1-\hwo+\eps.$$
Clearly, since in general this 4 equations can not be (nor we care for them
to be) satisfied simultaneously, we look for a condition that will produce a
simple (possibly) double
cusp, $\textit{both}$ at $N_e$ and $-N_e$. This would describe the location of
the two
broad peaks that are, in all cases,
at the limit between the "chaotic" and the "regular" region. For obtaining this
symmetric flat cusp
condition we pick the equation that is right in the middle of the set of 4 equations
given above. This equation is clearly 
\begin{equation} N_{e}=2v-1,\end{equation}
with the nice property that the $\eps$ and the $\wo$ dependence
drops out. This is a result that should work well whenever there is
a well defined "plateau" region, that is, for $v \geq \sqrt{\wo}$ when $\wo
\geq 1$, or $v\geq 1$ when $\wo<1$. 
As can be seen in Fig.6, for $\wo=1$ and $v\geq 1$, this equation seems to describe well the location of
the peaks in the spectrum of an eigenstate that separate the plateau from the
exponential decay region. For other values of $\wo$ it gives also good
agreement with the data.

To investigate further the structure of the FES of the system, we use Eq.(\ref{eigenvecfinal}),
and evaluate at $t=0$,
\begin{equation}
\begin{aligned}
|\phi^{-\eps}(0)\bra~&=~\ct |a\bra~+
~\st |b\bra,\\|\phi^{~\eps}(0)\bra ~&= -\st|a\bra ~+~ \ct|b\bra,
\end{aligned}\label{initialFstate}
\end{equation}
and where, using  $K(0)^2 + Z(0)^2 = 1$,  we have defined the angle $\theta$
so that $K(0) = \ct$, and $Z(0)=\st$. 

We now  concentrate on the study of the quantities $\ct(v),~\st(v),
~\theta(v)$, which, as we will see in the next sections, play an important part in
the evolution of the system, and correspondingly in the Dynamical Localization
and Harmonic Generation of the system. In what follows we will study them for three different values of the parameter
$\wo$:  Driving frequency above resonance, for $\wo=0.5$;  resonant case for
$\wo = 1$ and driving frequency below resonance, for $\wo=2$.

In Figs. 7,8,9, we show the results for $\wo =0.5$, $\wo = 1.0$ and  
$\wo = 2.0$ respectively. The upper panels show the dependence of the
 functions $\cttv$ and $\sttv$, and the lower panel
 the functions $\theta (v)$ and $\eps(v)$. As a function of the driving field
 amplitude $v$, these functions show an
oscillatory behavior, which can  be better appreciated in the 
lower panels, where
the function $\theta (v)$ is plotted along side with $\eps (v)$. An
interesting feature of this functions is the fact that in general, and to a
good approximation, 
there seems to be a phase shift of $\approx \pi/2$ 
between $\eps(v)$ and $\theta (v)$,  which 
corresponds to a distance of $\Delta v \approx 0.4$ between their
maxima.  This means
that to (almost) every maxima of $\theta (v)$ corresponds a zero of
 $\eps (v)$ (the DL
points) and vice versa, with the only exceptions occurring at the first
maxima or the first zero of this functions. Notice also that $\theta
(v)$ does not oscillate symmetrically around zero, with
the clear consequence that the
peaks in the functions $\cttv$ and $\sttv$ have
uneven heights, with alternating high and low peaks, although with
an overall decrease in height.
This alternating pattern of maxima seems to be
a general feature of the system, which we have found for all the values of 
 $\wo$
that we examined.

The DL regime, which can be characterized by the parameter $\alpha' \equiv
\wo~ \textrm{min}(1,1/\sqrt{v})<<1$, has been studied in several works. In this regime it
was found \cite{Dakhnovskii,Santana} that, at $t=0$, the eigenstates of the
system are of the form 
\begin{equation}
\begin{aligned}
|\phi^{-\eps}(0)\bra~&=~|a\bra~+
~\wo \frac{\pi}{4}H_0 (4v) |b\bra,\\|\phi^{~\eps}(0)\bra ~&= -\wo \frac{\pi}{4} H_0 (4v)|a\bra ~+~ |b\bra
\end{aligned}\label{DLregimeFES}
\end{equation}
where $H_0 (4v)$ is the Struve function\cite{Abramowitz} defined as
\begin{equation}
H_0(4v) \equiv \frac{4}{\pi} \sum_{n=0}^\infty \frac{J_{2n+1} (4v)}{2n+1} 
\end{equation}

In Figs. 7,8,9, in the lower panel we have plotted with a dash-dot line the
function $ \wo \frac{\pi}{4} H_0 (4v)$. As it was expected, since in the DL
regime $\st<<1$, then $\st\approx \theta$, and  $\theta(v) \rightarrow  \wo
\frac{\pi}{4} H_0 (4v)$. The convergence of $\theta(v)$ to the Struve function
is clearly slower for the bigger values of $\wo$. It is however rather
remarkable that the Struve function seems to approximate well the
function $\theta(v)$ much before the DL regime condition is satisfied
($\alpha\ll 1$);  
for example, for $\wo =0.5$, and $v=1$, or $\wo =1$ and $v=4$, which both give  $\alpha=0.5 \lesssim 1$  we already
see a good agreement between those two functions.
One can also see from Fig.7,8,9 that the Struve function correctly gives the
height of the maxima in $\theta (v)$, with the difference between those two
functions being mostly a phase difference that tends to zero as $v\rightarrow
\infty$. From the asymptotic form of the Struve function in the limit
$v\rightarrow\infty$ we obtain that, in that limit,  
\begin{equation}
\theta (v) \sim \wo
\frac{\pi}{4}\sqrt{\frac{1}{2\pi v}}\textrm{sin}(4v-\pi/4)\approx
\alpha~  \textrm{sin}(4v-\pi/4)  ~~~~~\textrm{for}~~ v\rightarrow\infty.\label{asympformtheta}
\end{equation}
where we defined $\alpha\equiv \sqrt{\pi/32}~ \alpha'=\sqrt{\pi/32}\frac{\wo}{\sqrt{v}}$, for $v>1$.
We will call $\alpha$ the "DL parameter". Eq. (\ref{asympformtheta}) establishes a nice connection between
the DL parameter and the amplitude of the oscillations in the function $\theta(v)$.

\subsection{Time-evolution operator}
To find out the time evolution of an initial state of the form
 $|a\bra$ or $|b\bra$, we can use the eigenstates
at time t=0, as given by Eq.(\ref{initialFstate}).
From that equation we can express $\ka$ and $\kb$ in the form
\begin{equation}
\begin{aligned}
|a\bra&=\ct|\phi^{-\eps}(0)\bra~-
~\st|\phi^{\eps}(0)\bra,\\|b\bra&= \st|\phi^{-\eps}(0)\bra
~+~ \ct|\phi^{\eps}(0)\bra.
\end{aligned}\label{ab(phi's0)}
\end{equation}
From these, it is now straight forward to construct two orthogonal
solutions to Schr\"{o}dinger's equation which, at $t=0$,
correspond to each one of the eigenstates of $H_0$ :
\begin{equation}
|a(t)\bra =e^{i\eps t}\ct |\phi^{-\eps}(t)\bra -
e^{-i\eps t}\st|\phi^{\eps}(t)\bra
 $$and$$
 |b(t)\bra=e^{i\eps
t}\st |\phi^{-\eps}(t)\bra +
e^{-i\eps t}\ct |\phi^{\eps}(t)\bra,
\label{a(t)b(t)}
\end{equation}
where $|a(0)\bra=|a\bra$ and $|b(0)\bra=|b\bra$. Using
Eq.(\ref{eigenvecfinal}) into above expressions we get

\begin{equation}
|a(t)\bra=[\ct K(t) e^{i\eps t}+ \st Z^* (t)e^{-i\eps t}]|a\bra
+ [\ct Z(t) e^{i\eps t}-  \st K^* (t)e^{-i\eps t}]|b\bra
 $$and$$
 |b(t)\bra=[\st  K(t) e^{i\eps t}- \ct Z^* (t)e^{-i\eps t}]|a\bra +
[\st Z(t) e^{i\eps t}+\ct K^* (t)e^{-i\eps t}]|b\bra .
\label{evolutioneq}
\end{equation}

The time-evolution operator for this
system, which is defined by the equation $|\Psi(t)\bra =
\textit{U}(t,0)\|\Psi(0)\bra$, can be extracted, in the basis $\{\ka,\kb\}$,
directly from the expressions above
\begin{equation}
\it{U_{ab} (t,0)} = \left[\begin{array}{rr}
                u_{aa}(t) &   -u_{ba} ^* (t) \\
                u_{ba}(t) & u_{aa} ^* (t)
            \end{array}\right],
\label{Uabmatrix}
\end{equation}
where,
\begin{equation}
u_{aa} (t)=\ct K(t) e^{i\eps t}+ \st Z^* (t)e^{-i\eps t},$$ $$
u_{ba} (t)=\ct Z(t) e^{i\eps t}- \st K^* (t)e^{-i\eps t}.
\label{defUs}
\end{equation}
The subindex $ab$ in $U_{ab}$ is there to remind us of the basis used to
represent the time-evolution operator.

Another way to arrive at the same result is by making use of the general 
expression [Eq.(27) in
Milfeld and Wyatt \cite{MilfeldWyatt}]
\begin{equation}
\textit{U}(t,0)=\Phi (t) e^{-i\Lambda t} \Phi^{-1}(0)
\end{equation}
where the columns of the Floquet matrix $\Phi (t)$ are the components of the
FES of the system, and $\Lambda$ is a diagonal matrix whose components are the
quasi-energies of the system.  For our system, from Eq.(\ref{eigenvecfinal}),
the Floquet matrix is

\begin{equation}
\Phi (t) = \left[\begin{array}{rr}
               K(t) &   -Z ^* (t) \\
                Z(t) & K^* (t)
            \end{array}\right],
\label{Floquetmatrix}
\end{equation}
and $\Lambda=-\eps \sigma_z$.

\section{Dynamic localization}
It is of particular interest to study the situation where the 
the system's initial state is a superposition of the eigenstates of $H_0$. In
particular, specially in the context of quantum wells, a particularly
interesting initial state is one in which the electron is localized in one of the wells. Such an initial state
corresponds to
$|\Psi (0)\bra =|1\bra \equiv \frac{1}{\sqrt{2}}(\ka + \kb)$, or $|\Psi (0)\bra
=|2\bra \equiv \frac{1}{\sqrt{2}}(\ka - \kb)$. In this case, we are
interested in calculating the probability to find the system at that same
location after a time $t$. The solution can be found by applying the
transformation $V=\frac{1}{\sqrt{2}}(\sigma_z + \sigma_x )$, which corresponds to the
change of basis $\{\ka,\kb\} \rightarrow \{|1\bra,|2\bra\}$, to the evolution
matrix in Eq.(\ref{Uabmatrix}). The resulting matrix, $ \textit{U}_{12}=V
\textit{U}_{ab}V^{-1}$, can be written as
\begin{equation}
\textit{U}_{12} (t,0) = \left[\begin{array}{rr}
               u_{11} (t) & - u_{21}^*  (t)  \\
               u_{21}  (t) &  u_{11} ^* (t)
 \end{array}\right],
\label{U12matrix}
\end{equation}
with
\begin{equation}
u_{11}=Re(u_{aa})+iIm(u_{ba}),$$ $$
u_{21}=Re(u_{ba})-iIm(u_{aa}). \label{U12def}
\end{equation}

It is a well-known fact that dynamic localization occurs only at the points where
the quasi-energies vanish, i.e. $\eps (v) = 0$. At this particular values of driving
field amplitude $v$, as it can be
seen from Eqs.(\ref{Uabmatrix}) and (\ref{defUs}), the dynamics of the system
becomes strictly periodic, with period $2\pi/\omega$, for any initial
condition.
The points where $\st (v)=0$ also
produce a  periodic evolution of the system (except for an overall time-dependent phase factor), 
but only for the particular initial conditions  $|\Psi (0)\bra=\ka$ or
 $|\Psi (0)\bra =\kb $.

To study the system at the DL points, we will look at the quantity $|u_{11}
(t) |^2$, which gives the
probability (as a function of time), that the system, if localized in
a well (e.g. $|1\bra$ ) at time $t=0$, would remain in that well at a later
time t. For $\eps (v)=0$, and from Eqs.(\ref{defUs},\ref{U12def}) we get
\begin{equation}
u_{1,1} (t) = \ct [Re(K(t)) + iIm(Z(t))] + \st [Re(Z(t)) +
iIm(K(t))]~,~~~~~ \rm{at~ DL~ points}.
\end{equation}
Since this probability is periodic, $u_{1,1} (0)=u_{1,1}
(2\pi/\omega)=1$, it is interesting to see what is the value of this function
half-way 
through a cycle of the field, that is, for $\omega t=\pi$. From Eq.(\ref{KtZt})
we can
deduce that $ K(\pi/\omega)=K(0)=\ct$, and $Z(\pi/\omega)= ~-Z(0)=-\st$.  Using this, we get from the
above equation 
\begin{equation}
u_{1,1} (\pi/\omega) = \ct^2-\st^2 = \ctt  ~,~~~~ \rm{at~ DL~ points}.
\label{degreeofDL}
\end{equation} 

We will show in the following figures that the function $\ctt$  plays
an important role in quantifying how good is the localization of the system
at the DL points. 

In the lower panels of Figs.7,8,9, we have placed dots on the quasi-energy
functions $\eps (v)$, at different values of
$v$. In the upper panel in Figs. 10,11,12 we show, for each one of those
points,  the corresponding probability
$|u_{1,1} (t)|^2$, as given by Eqs. (\ref{defUs}),(\ref{U12def}).
This probability gives the
 location of the particle, as a function of time, when starting from the
 initial state $|1\bra$.
In Fig.10 we study the case $\wo = 0.5$. In this plot one can see that the probability evolves in a manner that is typical of the high frequency
 regime which has been amply studied in the literature. The function can be
 described as of the form $cos(\eps t)$+ small time-periodic oscillations. 
Clearly, at the DL points (A,B,C,D), when $\eps =0$, the probability is exactly periodic (period
$2\pi\omega$) and  oscillates between 1 and a value that depends on the driving
field amplitude $v$ and on $\wo$. As derived before, in the middle of the
driving period, the probability $|u_{1,1}|^2 = \cttv$. 

Given the correspondence between the maxima of $\theta(v)$ and the zeros of $\eps(v)$, we can
conclude that the minima of the function $\cttv$ tells us the value of the probability $|u_{1,1}(t)|^2$  half way through the driving field 
period ($t=\pi /  \omega$), at the corresponding DL point (with the exception
of point A). This value of  $|u_{1,1}(t)|^2$  is very important because it gives us
information about the extend to which the particle is localized when the DL points are reached. As we
can see in the lower panels of Figs. 10,11,12, when the function
$|u_{1,1}(t)|^2$ has a local minimum at $t=\pi / \omega$, then $\cttv$
is close to be the absolute minimum value that $|u_{1,1}(t)|^2$ reaches
through out a whole period, $P_{l} =min(|u_{1,1}(t)|^2)$.  It is therefore an
 important number for describing the
amount of localization that can be achieved in the system. When the function
$|u_{1,1}(t)|^2$ has a local maximum at $t=\pi / \omega$, then
$\cttv$ is still equal to that value but this
is not necessarily close to $P_l$ (specially when we are far from the DL regime). This is the case 
at points B and D in
those figures. Surprisingly, however,  $P_l$, for cases B and D, is given, to a certain approximation, by the value of the function $\cttv$ at the minima that corresponds to
the \textit{next} DL point. We do not know why this should be so but it was found
to be the case in all the cases that we examined.  According to this, the DL 
points can be grouped in pairs, with (B,C) having a similar amount of
localization, given by  $\textrm{cos}^2 (2\theta(v_c ))$; points (D,E) with a localization given
by $\textrm{cos}^2 (2\theta(v_E ))$, and so on. It is therefore the deeper minima in $\cttv$ that give us directly the amount of localization in the system. 

An interesting question to ask, because of practical applications, is whether there is a simple
relationship between the amount of localization $P_l$, $\wo$ and $v$. In
other words, given a particular amount of localization $P_l$, and a given
$\wo$, what is the required driving field amplitude for the system to achieve such level
of localization? To answer this question
 in the DL regime (were we expect a high level of localization, $P_l
 \lesssim 1$), one can use Eq.(\ref{asympformtheta}) to derive an approximate relationship between
$P_l$, $\wo$ and $v$. At the maxima of
$\theta(v)$ (Dl points),  $\theta (v) \sim \alpha$,  which gives
$P_l = \ctt \approx 1-4\theta^2=1- \frac{\pi}{8} \frac{\wo^2}{v}$, from
which one finally gets

\begin{equation}
 v \approx  \frac{\pi \wo ^2}{8(1-P_{l} )}.
\end{equation}

This simple equation, which has been derived assuming $v>1,~\alpha<<1$,
already gives good results even for a localization as low as $P_l =0.5$, with
the accuracy improving further for higher values of $P_l$. For $P_l =0.5 $ the result
is $v\sim 0.78~ \wo ^2$, for $P_l = 0.9$ one obtains  
$v\sim 4~ \wo ^2$ and for $P_l = 0.99$ it gives $v\sim 40~ \wo ^2$.


\section{High-order Harmonic Generation}

The expected value of the dipole moment of the system can be calculated from the expression
\begin{equation}
\ket d(t) \bra = \ket \Psi (t) | X|\Psi (t) \bra =\mu
\ket \Psi(0)| \sigma_{x} (t)|\Psi(0)\bra~,
\end{equation}
where
\begin{equation}
\sigma_{x} (t) \equiv U_{ab}(t)^\dagger \sigma_{x} U_{ab} (t),
\end{equation}
and using Eqs.(\ref{Uabmatrix}), we get
\begin{equation}
\sigma_{x} (t) = \left[\begin{array}{rr}
              Re[ u_{aa}^*  (t)  u_{ba} (t)] &  [u_{aa}^2  (t)- u_{ba}^2  (t)]^*  \\
         u_{aa}^2  (t)- u_{ba}^2  (t)   &   -Re[ u_{aa}^*  (t)  u_{ba} (t)] .
 \end{array}\right].
\label{sigmaxt}
\end{equation}
For a general initial condition of the form $|\Psi(0)\bra = c_a \ka + c_b \kb$
we get
\begin{equation}
\ket d(t) \bra/ \mu  = Re[u_{aa}^* (t) u_{ba}(t)](|c_a|^2-|c_b|^2) + 2 Re[c_a c_b
  ^*(u_{aa}^2 (t)-u_{ba}^2 (t))].\label{generaldipole}
\end{equation}
If we now define the following time-periodic functions,
\begin{equation}
\begin{aligned}
R(t) & \equiv Re(K(t))Re(Z(t)) + Im(K(t))Im(Z(t)) &\textrm{(odd harmonics),}\\
Q(t) & \equiv Re(K(t))Im(K(t)) - Re(Z(t))Im(Z(t)) &\textrm{(even
  harmonics),}\\
P(t) & \equiv Im(K(t))^2 + Re(Z(t))^2 &\textrm{(even harmonics),}
\end{aligned}\label{RQP}
\end{equation}
then, using this, and plugging Eqs.(\ref{defUs}) into Eq.(\ref{generaldipole}), we obtain (after some lengthy algebra) the
final expression
\begin{equation}
\begin{aligned}
\ket d(t) \bra/ \mu & = \left[ - \stt \ce + 2\ctt R(t) +2 \stt \se Q(t) + 2\stt \ce
P(t)\right] \opt \\
&+ \left[  \ctt \ce + 2 \stt R(t) -2 \ctt\se Q(t) -2 \ctt\ce P(t)\right]\Rtun \\
& + \left[ -  \se - 2 \ce Q(t) + 2\se P(t)\right] \Ctun ~.
\end{aligned}\label{masterdipole}
\end{equation}
The quantities $\theta$, $\eps$, $P(t),~Q(t),~R(t)$
depend on $v$, even though, we do not explicitly indicate so in this
notation.
We will refer
to the initial condition where $|c_{a}|=1$ or  $|c_{b}|=1$ (and therefore
$c_{a}c_{b}^* = ~0$), as 
the "Optical" Initial Condition (OIC); the case $c_a = \pm c_b$ as the
"Tunneling" initial condition (TIC) and the case $c_a = \pm i c_b$ as the
"Complex Tunneling" Initial Condition (CTIN).

From the expectation value of the dipole moment, one can obtain the
(coherent) emissions spectrum of the system from the expression
$|d(\Omega)|^2=\left| \frac{1}{T} \int_{t_0}^{t_0 +T}dte^{i\Omega t} \ket
d(t)\bra \right|^2$.

Eq.(\ref{masterdipole}) deserves many comments. Notice that generally, there is low-frequency generation, with frequency $2\eps$ for all initial conditions. Only
at some particular driving amplitudes this can be prevented. For OIC the low-frequency component can be suppressed
only when $\stt=0$; for TIC only when $\ctt=0$. There
is always low-frequency generation at any driving amplitude for CTIC.
There is odd-harmonics generation, at frequencies $(2n-1)\omega$ and due to the
terms with $R(t)$. This harmonics are generated except when
$\ctt=0$ for OIC and except when $\stt=0$ for TIC. Interestingly
enough, CTIN\textit{can  not generate odd harmonics}. 
Even harmonics, at frequencies $2n\omega\pm 2\eps$ occur because of the
presence of terms $(\se$ or $ \ce) Q(t)$ and $(\se$ or  $\ce )P(t)$, and are generated except when
sin$2\theta(v)=0$ for OIC and except when
cos$2\theta(v)=0$ for TIC. They are always present for $CTIC$.

For a general driven Hamiltonian that possesses the Dynamical Symmetry (DS)
$H(x,t)=H(-x,t+\tau /2)$ it has been proven \cite{Moiseyev} that the HHG from a Floquet eigenstate consists of only
\textit{odd}-harmonics. Our Hamiltonian, Eq. (\ref{2Sham}), is certainly of
this kind since the quantum
operator that corresponds to $x$ is $\sigma_x$, and therefore, if $\sigma_x
\rightarrow -\sigma_x$, and $t\rightarrow t+\tau/2$, the Hamiltonian in
Eq. (\ref{2Sham}) remains the same. According to this, a good check for our
result can be done by replacing in Eq.(\ref{masterdipole}) the initial
condition $c_a=\ct$, $c_b=\st$ which corresponds
to the initial state being already a Floquet eigenstate [see
  Eq.(\ref{initialFstate})]. After doing that, one
obtains $\ket d(t)\bra/ \mu = 2R(t)$, which means that this state only generates odd-harmonics, in agreement with the dynamical symmetry arguments
developed in the citation above.

As our result (\ref{masterdipole}) shows, in a two-level system it is possible
to generate both even \textit{and} odd harmonics. This has also been found in
numerical calculations for a double quantum well \cite{Bavli}.  The kind of harmonics present
in the spectrum is dependent on the initial condition, the driving
amplitude and also the phase of the driving field. In this work we have chosen
a driving force of the form $\textrm{cos}(\omega t)$; however, in the most
general case of time dependence of the form $\textrm{cos}(\omega t+\phi)$, the driving phase $\phi$, in general,
plays an important role in determining  the harmonic content of the spectrum.
For the case $\textrm{sin}(\omega t)$ see Gauthey et al\cite{Gauthey95}.

It has been argued\cite{Moiseyev} that in systems  that get ionized during
the course of the interaction with the field, only one
resonant Floquet state (the one with the longest life-time) contributes to the
harmonic spectrum and therefore this could
explain the absence of even-harmonic generation in most experiments. 
There are several experimental results
that involve lasers interacting with solid targets, in which even-harmonics
and odd harmonics are generated. Even though ionization takes place, the presence of
even harmonics has been attributed to the lack of DS in the
Hamiltonian. 
For systems where the main source of harmonic
radiation are bound-bound transitions, it is not clear how to justify the use of only one of the FES 
in the calculation of the HG of the system. In those kinds of system we will argue in future work that, if the evolution of the system is adiabatic under the effect of a laser pulse, then again, only odd-harmonics are generated. For non-adiabatic evolution, we expect to see generation of even and odd harmonics and also dependence on the initial phase of the driving.
It is interesting to check our results for the time dependent dipole moment
of the system in the DL regime. As we showed before, in the limit of perfect
DL, where $v\rightarrow \infty$ or $\wo \rightarrow 0$, one gets
$\theta=0$, $K(t)$ becomes pure real, $Z(t)$ becomes pure imaginary, and therefore
$R(t)=Q(t)=P(t)=0$. The result for the dipole moment is therefore
\begin{equation}
\ket d(t)\bra/ \mu = \ce \Rtun - \se \Ctun = 2 |c_a c_b|\textrm{cos}(2\eps t + \gamma),
\end{equation}
with $\gamma$ being the phase difference between $c_a$ and $c_b$. This is the
zeroth-order expression in the first-order perturbative results obtained by
Delgado and Gomez-Llorente \cite{Delgado}. Clearly, in this limit there is
\textit{no} HG. From this it is clear that in the case of a driven two-level
system, DL and HHG are at odds; the greater the localization, the smaller the
HHG. The amplitude of any harmonic component of the emission spectrum of the
system (except
the low frequency one), as a function of increasing driving field strength $v$  should
initially increase, then plateau (with oscillations) and finally decrease as
$v$ is further increased.

Near the DL regime we can obtain some expressions and 
compare them with the perturbative results obtained in other works
\cite{Delgado,Ivanov,Dakhnovskii}.
For $v>1$ and $\alpha=\sqrt{\frac{\pi}{32}}\frac{\wo}{\sqrt{v}}$, we found that the behavior of the functions $K(t)$ and $Z(t)$ is
\begin{equation}\begin{aligned}
K(t) &\approx Re(K(t)) + i\alpha \kappa(t)  \\
Z(t) &\approx \alpha \zeta (t)  + i Im(Z(t)).
\end{aligned}
\end{equation}
Here $\kappa (t)$  and $\zeta (t)$ are periodic functions (even and odd respectively) with amplitude close to one.
As we have shown before, in this regime, the amplitude of the oscillations in
the function $\theta(v)$ is also proportional to $\alpha$.
According to this, and using Eq.(\ref{RQP}), we see that in this regime, 
$R(t) \propto \alpha$, $Q(t) \propto \alpha$, $P(t) \propto \alpha^2$. We now
rewrite Eq.(\ref{masterdipole}) in a form that makes evident the magnitude of
the
different terms that contribute to the dipole moment,
\begin{equation}
\begin{aligned}
\ket d(t) \bra & \approx  \left[ - \alpha\tilde{\theta} \ce +  \alpha \tilde{R}(t) +
  2\alpha^2 \tilde{\theta} \se \tilde{Q}(t) + 2\alpha^3 \tilde{\theta} \ce
\tilde{P}(t)\right] \opt \\
&+ \left[  \ce  -2 \alpha \se
  \tilde{Q}(t)+ 4\alpha^2 \tilde{\theta} \tilde{ R}(t)  -2 \alpha^2 \ce \tilde{P}(t)\right]\Rtun \\
& + \left[ -  \se - 2 \alpha \ce \tilde{Q}(t) + 2\alpha^2 \se \tilde{P}(t)\right] \Ctun ~.
\end{aligned}\label{DLdipole}
\end{equation}
where $\tilde{R}(t)= R(t)/\alpha$, $\tilde{Q}(t)= Q(t)/\alpha$, $\tilde{P}(t)=
P(t)/\alpha^2$, and $\tilde{\theta}= \theta / \alpha \sim \textrm{sin}(4v-\pi/4)$. 
From this expression we can see that in the DL regime, OIC generates low
frequency and odd-harmonic radiation to second order in $\alpha$ and even harmonic
generation is forth order in this parameter. For TIC there is strong low
frequency generation independent of any of the parameters of the
system (provided we are in DL regime
); even harmonic generation is second order
and odd harmonic generation is forth order in $\alpha$. For CTIC
the situation is similar to TIC except that no odd-harmonic radiation is
generated at all. 

As mentioned before, from the general expression for the dipole moment, since
 $\eps=\eps(v)$ and $\theta=\theta(v)$, with successive points where either
$\theta(v)$ or $\eps(v)$ vanish, there is a lot of variation in the
composition (in terms of odd or even harmonics) of the spectrum of this
system as a function of the amplitude of the driving field. The interesting
exception being the CTIC where even harmonics are always generated and odd
harmonics are never generated.

In Fig.13, 14 and 15 we show the emission spectrum for  three
distinct initial conditions (OIC,TIC,CTIC) and for two values of the driving
field amplitude. What is common to all this plots is the typical profile of
harmonic generation that has been found theoretically and experimentally in
different systems. It consists of a plateau where the amplitudes are of
similar magnitude (fluctuations are of one to two orders of magnitude typically),
followed by a frequency cut-off after which the harmonic amplitudes decay very
quickly (exponentially in $n\omega$).

From the structure of the FES, like the one shown in Fig. 6, one can see that there is a cut-off
in the harmonic components of the
eigenstates, which occurs at $N_{e}=2v-1$ (for $ v\geq 1$). The location of the cut-off in the emission spectrum of the system follows from the
location of the cut-off in the eigenstates, since the functions $R(t),Q(t),P(t)$ are second order in the FES
components. $R(t), Q(t),
P(t)$ therefore have cut-offs at $N_{cut}=2N_{e} =4v-2$. This is the same cut-off that we expect to see in the emission spectrum of the system. 

As we can see in Fig. 13, 14, 15, this result describes quite well the location of the last
peak before the components start to decay exponentially. 
The linear dependence of $N_{cut}$ on the amplitude of the field implies that
$N_{cut} \propto \sqrt{I}$, where I is the intensity of the field. This result
 is expected for HHG in systems where bound-bound transitions are
 dominant \cite{Averbukh,Zdanska}. For dominant bound-continuum
 transitions \cite{Corkummultiple}, $N_{cut} \propto I$. 

In Fig.13 we show the emission spectrum for OIC, and for two different
values of the driving amplitude. For $v=4.15$ (which gives $\theta\approx 0$),
we see, from the first line of Eq.(\ref{DLdipole})that one expects no low
frequency component, odd harmonics and no even harmonics. For $v=4.5$ (which gives $\eps \approx 0$), we obtain a strong low
frequency component, strong odd-harmonic components and very weak even-harmonics
components that are due to the function $P(t)$ only (since at $\eps = 0$, 
 $\se =0$ and therefore  $Q(t)$ does not contribute to the spectrum). This is
a special case since in general, in this regime, $Q(t)$'s contribution to
$|d(\Omega)|^2$ is two orders of magnitude bigger than the one from $P(t)$.

In Fig. 14 we show the emissions spectrum for TIC for the same values of
$v$ as before. In the $\theta = 0$ case, we can see, from
the second line in Eq.(\ref{DLdipole}), that there is a strong low frequency
component ($\Omega=\pm 2\eps$), no odd-harmonics  and split even harmonics ($\Omega=2n\pm2\eps$). For the case $\eps=0$
there is a zero frequency component and also odd and even harmonics (not split) of the
same magnitude.

Finally, in Fig. 15 we show the case when the initial condition is
CTIC. From the third line in Eq.(\ref{DLdipole}) we see that, for the case
$\theta=0$ we should get a strong low frequency component and split even
harmonics. For $\eps=0$ case there is no static component, only even
harmonics (not split).

For the characteristic plateau of HHG to appear, it was shown \cite{Kaplan}
that $v\geq \sqrt{\wo}$. In this region we have found numerically that $\ctt
\neq 0$,  which means that in the OIC, overall, the odd harmonics are much more stable
with respect to changes in the driving amplitude that the
even harmonics, which as a whole, do fluctuate with $v$ (following the behavior of
$\stt$). As one approaches the DL regime the odd harmonics will dominate the
spectrum for the OIC (which is the initial condition most easily obtained
experimentally). It is easy to derive
an estimate for the conditions under which  odd-harmonics will  dominate over even
harmonics in the OIC.  For that we take $\alpha \geq 0.1 $, so that there is
at least one order of
magnitude difference in their contributions to Eq.(49). From  $\alpha =
\sqrt{\frac{\pi}{32}}\frac{\wo}{\sqrt{v}}$ one gets the estimate
$$ v \gtrsim 10\wo^2 .$$
When this condition is satisfied, odd harmonics will clearly dominate the spectrum in the OIC.

\section{Summary and Conclusions}
In this work we used a complete Floquet-Green operator formalism for the
analytical solution of a harmonically driven two level system. From this operator we were
able to completely solve the system and obtain the quasi-energies and the
eigenstates in terms of continued fractions.

We found that the oscillatory behavior of the quasi-energies of
the system can be interpreted as due to multi-photon resonances that occur in
an increasingly periodic way as the strength of the driving field is increased. This
multi-photon resonances involve an odd number of photons and produce avoided
crossings centered at the edges of the energy Brillouin zone $\eps=\pm
0.5$. The width and amplitude of them increases with the driving field
amplitude. The structure of the
continued fractions from which the quasi-energies can be obtained made it clear
that this is the case, since each term in the continued fraction was found to
be related to the inclusion of a particular basis state in the calculations,
and each avoided crossing studied did only appear when the corresponding term
in the continued fraction was included.

We were able to obtain analytical expressions for the components of the
Floquet eigenstates and from them we constructed the time-evolution operator of
this system.  The structure of the eigenstates revealed a characteristic
pattern, with a "chaotic" region where the components are sensitive to small
changes in $v$, change sign and have
zeros, and a "regular" region where the components depend less sensitively on
$v$, and decay exponentially with the
frequency $n\omega$. The cut-off frequency that divides the two regions and
gives the position of the largest floquet component of the eigenstates
is proportional to the driving field strength $v$, $N_{e} = 2v-1$, a result that is valid for
any $\wo$ (provided $v > \sqrt{\wo}$). 

From the time evolution operator we were able to study the Dynamic Localization
phenomena which is one of the important features of this
Hamiltonian. It is known that DL is obtained at values of the driving field strength for which
the quasi-energies vanish. In Quantum Computation, this localization mechanism
can be of great importance for 
controlling the state of a q-bit (as realized for example in a solid state
devise such  as a double quantum well or a Josephson Junction), since the amount of tunneling of an electron between two adjacent
wells can be varied with the intensity of the driving field. 
In this work we obtained a correlation between the DL
points and the maxima of the quantity $\theta (v)$ that describes the Floquet
eigenstates of the system. Also, the value of the function $\cttv$,
at its lowest minima,
was found to give a good approximation to the degree of localization of the
system at the corresponding DL points.  These points appear in pairs with
simmilar degrees of localization corresponding to each pair. 
We also found the equation  $$v \approx  \frac{\pi \wo ^2}{8(1-P_{l}
  )},$$ which, at the DL points, relates the degree of localization $P_l$, with the
 energy level difference $\omega_0$, and the amplitude $v$ of the
driving field (all quantities in units of $\hbar\omega $). This equation can be used to
easily estimate the amount of localization achievable in this system for a given set of parameters. 

For the study of  High Harmonic Generation in this system we calculated the
expected value of the dipole moment which revealed that the emission
spectrum, in general, can contain a low frequency component and odd and
even harmonics. The particular components and their amplitudes depend strongly
on the initial state of the system and on the specific driving field
amplitude. We found that there exists an initial condition (Complex Tunneling) of the driven
two-level system for which no odd-harmonics can be generated at any driving
field strength. For each one of the "pure" initial conditions, namely Optical (OIC),
Tunneling (TIC) or Complex Tunneling (CTIC), the driving amplitudes where the $\theta(v) =0$
produce either pure even harmonic generation or pure odd harmonic generation,
depending on the particular initial condition. For all other driving
amplitudes there will be in general generation of both kinds of harmonics
(except for CT initial condition). When the initial state is a Floquet
eigenstate, we found that only odd-harmonics are generated, a result that is
expected from the Dynamical Symmetry (DS) of the Hamiltonian.
In general we can say that the emission spectrum of the driven two-level
system is sensitive to different parameters, such as the initial preparation, the amplitude
of the driving and  its phase (phase dependence will be explored in
future work). 

For small values of $\alpha$, but still not deep into the DL regime ($\alpha
\approx 0.15$)  we showed the different magnitude of the terms that give low frequency, even harmonics and odd harmonics. In the Optical Initial Condition (OIC) we found that odd harmonic generation is stronger, with even harmonics weaker and even nonexistent for particular values of the driving field amplitude (when $\theta (v) = 0$).  For Tunneling Initial Condition (TIC), even harmonics dominate, except at the DL points ($\eps =0$), where even and odd harmonics have approximately the same amplitude. For Complex Tunneling Initial Condition (CTIC) no odd  harmonics are generated.

Well inside the DL regime, odd harmonics are almost exclusively generated in
the Optical Initial Condition, and even harmonics in both of the Tunneling
Initial Conditions. In this regime of high localization, Harmonic generation  decreases  with decreasing $\alpha$, a situation that clearly places both effects on opposite sides of the parameter
space, although with significant overlap between them. In general, for a driven two level system, the strongest HHG should occur for higher values of $\wo$ and for driving amplitudes
$v \lesssim 10 \wo^2$ ( for $\alpha > 0.1$). For strong DL, smaller values of $\wo$ and/or bigger values of $v$ are required, $v \gtrsim 10 \wo^2$ (for $\alpha < 0.1$). \textit{Strong} HHG in a
two-level system is an eminently non-linear, non-perturbative phenomenon.

The driven two-level model has been shown to be useful
in understanding the basic features of high-harmonic generation experimentally
observed in atoms and molecules, even though obviously the full complexity of
the general
processes in atoms and molecules can
not be accounted for with this model.  The model predicts an
emission spectrum with a plateau and a cut-off frequency ($N_{cut}= 4v-2$),
whose location
depends on the square-root of the intensity of the driving field (since $v\propto E
\propto  \sqrt{I}$). Such dependence has been found in systems where
bound-bound transitions account for most of the HHG \cite{Averbukh,Gupta}. 
For the case of charge-resonant states of
odd-charge molecular ions it has also been argued\cite{Ivanov} that this
simple two-level model can account for the behavior of these molecules in a
strong laser pulse. A square-root dependence of the cut-off frequency with the intensity of the laser field can be found in
several experiments\cite{L'Huillier}, a fact that is characteristic of a
driven two-level system and which might point in the direction of an effective two-level dynamics for some systems \cite{Kaplan}. 

HHG and DL are the two mayor features of a driven two-level system with a
great range of applications in atomic and solid state systems where
coherent sources of radiation and control of quantum states are
sought. Through this work we have established a strong connection between this
two phenomena and provided a detailed description and a deeper understanding of this
system.


\section*{Acknowledgments}
The Author would like to thank Prof.  L.E. Reichl for many years of advise and
support. This work was partly funded by the The Robert A.
Welch Foundation (Grant No. F-1051) Êand the Engineering Research 
Program Êof the Office of Basic Energy Sciences at the U.S. 
Department of Energy (Grant No. DE-FG03-94ER14465).


\section*{Appendix A: components of the FES}

From the definition of the Floquet-Green operator, we have shown
in Eq.(\ref{G00}) that the residue of the function $\G_{0,0}(E)$
at a pole $E =\eps +p$ is the operator (2x2 matrix)
$|\phi_{p}^{\eps}\bra\ket\phi_{p}^{\eps}|$, where
$|\phi_{p}^{\eps}\bra$ is the $pth$ Fourier component of the
Floquet eigenstate corresponding to the quasi-energy $\eps$. If we
write $|\phi_{p}^{\eps}\bra\equiv K_p|a\bra+ Z_p|b\bra$, then,
in principle, from the residue (a matrix) of $\G_{0,0}$ we could
determine the components $K_p,~Z_p$.  This is only
true, however, provided the residue does not turn out to be a
diagonal matrix. Unfortunately, this is the case for the system we
are considering, since, as we mentioned before, $G_{0,0}(E)$
\textit{is} a diagonal matrix ($V_{eff}(E)$ is diagonal) and
therefore, its residue only gives us information about $|K_p|^2$
and $|Z_p|^2$. From Eq.(\ref{Gandeigenveccomp1}) we can see that
the off-diagonal components $G_{m,0}$ can give us the needed
components:
\begin{equation}
\G_{m,0} (E)=\sum_{\gamma,p}
\frac{1}{E-\eps_{\gamma}-p}|\phi^{\eps_{\gamma}}_{m+p}\bra\ket\phi^{\eps_\gamma}_{p}|~
.
\label{Gandeigenvecpzero}
\end{equation}
From this we get that at the pole $E=\eps _\gamma$ (and therefore $p=0$), the 
residue is
$|\phi^{\eps_{\gamma}}_{m}\bra\ket\phi^{\eps_\gamma}_{0}|$. From
all these residues for all values of $m$, one could therefore extract all the eigenvector
components $|\phi^{\eps_{\gamma}}_{m}\bra$.

The $\G_{m,0}$ operators can be obtained from $\G_{0,0}$, as it was shown in
Ref. \cite{Martinez03}. The result is

\begin{equation}
\begin{aligned}
\G_{1,0}&=F^\up _0 \G_{0,0}~,\\
\G_{2,0}&=F^\up _1 \G_{1,0}=F^\up _1 F^\up _0  \G_{0,0}~, \\
\vdots\\
\G_{m,0}&=F^\up _{m-1} \G_{m-1,0}= \prod_{j=0}^{m-1}F^\up _j 
\G_{0,0}(E)~,~~~~~~~~~~~\rm{ for~m>0},
\end{aligned}\label{Fupsimple}
\end{equation}
and where (showing explicitly the energy dependence)
\begin{equation}
F^\up _j (E)\equiv (V\F_{j,0}(E))^{-1}V =V^{-1}V^\up_{\rm{eff}}(E+j)\\
=\left[
\begin{array}{rr}
                 0~~~~ & \frac{1}{v}v_b ^\up (E+j)   \\
               \frac{1}{v}v_a ^\up (E+j)  &  0~~~~
            \end{array}\right]~,
            \label{upstairsoper}
\end{equation}
and similarly,
\begin{equation}
\G_{-m,0}(E)= \F^\down _{-m+1} G_{-m+1,0}=
\prod_{j=0}^{m-1}\F^\down _{-j}  G_{0,0}~~~~~~~~~~~\rm{
for~m>0},
\label{Fdownsimple}
\end{equation}
with
\begin{equation}
F^\down _{-j} (E)\equiv (V\F_{-j,0}(E))^{-1}V =V^{-1}V^\down_{\rm{eff}}(E-j)\\
=\left[
\begin{array}{rr}
                 0~~~~ & \frac{1}{v}v_b ^\down (E-j)   \\
               \frac{1}{v}v_a ^\down (E-j)  &  0~~~~
            \end{array}\right]~.\label{downstairsoper}
\end{equation}

It is not difficult to show using Eqs.(\ref{Gandeigenvecpzero}),
(\ref{Fupsimple}), and (\ref{Fdownsimple}), that

\begin{equation}
|\phi^{\eps_{\gamma}}_{m+1}\bra=F^\up _m
(\eps_\gamma)|\phi^{\eps_{\gamma}}_{m}\bra\\
$$\rm{and} $$
 |\phi^{\eps_{\gamma}}_{m-1}\bra=F^\down _m
(\eps_\gamma)|\phi^{\eps_{\gamma}}_{m}\bra~,
\label{stairelations}
\end{equation}
with $\eps_\gamma = \pm\eps$.

As we already mentioned, at $E=\eps$ the residue of $G_{0,0}(E)$
is proportional to $|b\bra$ and at $E=-\eps$ the residue is
proportional to $|a\bra$. From this, and using
Eq.(\ref{stairelations}) together with Eqs.(\ref{upstairsoper})
and (\ref{downstairsoper}), we can write explicitly the two
Floquet eigenstates of this system:

\begin{equation}
|\phi^{-\eps}(t)\bra=N[ ....+ ~\frac{e^{2i\w
t}}{v^2}v_b^\down(-\eps -1)v_a^\down(-\eps)|a\bra ~+ ~\frac{e^{i\w
t}}{v}v_a^\down(-\eps)|b\bra~+ ~|a\bra + $$ $$
 + ~\frac{e^{-i\w t}}{v}
v_a^\up(-\eps)|b\bra~+ ~ \frac{e^{-2i\w t}}{v^2}v_b^\up(-\eps
+1)v_a^\up(-\eps)|a\bra~+ .....]
$$and$$
|\phi^{\eps}(t)\bra=N[ ....+ ~\frac{e^{2i\w t}}{v^2}v_a^\down(\eps
-1)v_b^\down(\eps)|b\bra ~+ ~\frac{e^{i\w
t}}{v}v_b^\down(\eps)|a\bra~+ ~|b\bra ~+$$ $$
 + ~\frac{e^{-i\w t}}{v}
v_b^\up(\eps)|a\bra~+ ~ \frac{e^{-2i\w t}}{v^2}v_a^\up(\eps
+1)v_b^\up(\eps)|b\bra~+ .....],
\end{equation}
with N being a normalization constant. If we use the symmetry
relations, $v_b ^\down(-E)=-v_a^\up(E)$ and $v_b
^\up(-E)=-v_a^\down(E)$ we can write this two eigenstates in a
more compact way,
\begin{equation}
|\phi^{-\eps}(t)\bra=K(t)|a\bra~+ ~Z(t)|b\bra
$$and$$
|\phi^{\eps}(t)\bra= -Z(t)^* |a\bra ~+~ K(t)^* |b\bra ~,
\end{equation}
with
\begin{equation}
\begin{aligned}
K(t)&= N(....+ \frac{e^{4i\w t}}{v^4}v_b^\down(-\eps
-3)v_a^\down(-\eps -2)v_b^\down(-\eps
-1)v_a^\down(-\eps)~+~\frac{e^{2i\w t}}{v^2}v_b^\down(-\eps
-1)v_a^\down(-\eps) +1 \\ &+ \frac{e^{-2i\w t}}{v^2}v_b^\up(-\eps
+1)v_a^\up(-\eps)~+~\frac{e^{-4i\w t}}{v^4}v_b^\up(-\eps
+3)v_a^\up(-\eps+2)v_b^\up(-\eps +1)v_a^\up(-\eps)) \\
 &=N[\sum_{n=-\infty}^\infty e^{-2in\w t}K_{2n} (-\eps)]~,
\\
Z(t)&= N[...\frac{e^{3i\w t}}{v^3}v_a^\down(-\eps
-2)v_b^\down(\eps -1)v_a^\down(-\eps)+~\frac{e^{i\w
t}}{v}v_a^\down(-\eps) +\frac{e^{-i\w t}}{v}
v_a^\up(-\eps)~+~\frac{e^{-3i\w
t}}{v^3}v_a^\up(-\eps +2)v_b^\up(-\eps +1)v_a^\up(-\eps)....]\\
&=N[\sum_{n=-\infty}^\infty e^{-i(2n-1)\w t}Z_{2n-1}(-\eps) ]
\end{aligned}
\label{KtZt}
\end{equation}
where $N$ is a normalization constant so that $|K(t)|^2 + |Z(t)|^2=1$, and we have 
\begin{equation}
\begin{aligned}
K_{2n} (-\eps) &= \begin{cases}
1 & \text{for $n=0$}, \\ 
\prod_{j=1} ^{|n|} \frac{1}{v^2}v_b
 ^{sg(n)} (-\eps +sg(n)(2j-1)) v_a
 ^{sg(n)} (-\eps +sg(n)(2j-2))) & \text{for $ |n|>0$}
\end{cases}\\
Z_{2n+1} (-\eps) &= \begin{cases}
 \frac{1}{v}v_a^\up (-\eps +2n)K_{2n}(-\eps) & \text{for $2n+1>0$}, \\ 
 \frac{1}{v}v_a^\down (-\eps +2n+2)K_{2n+2}(-\eps) & \text{for $2n+1<0$},
\end{cases}
\end{aligned}\label{components}
\end{equation}
and we used $sg(x)\equiv x/|x|$.

\newpage

\begin{figure}
\begin{center}
\epsfig{figure=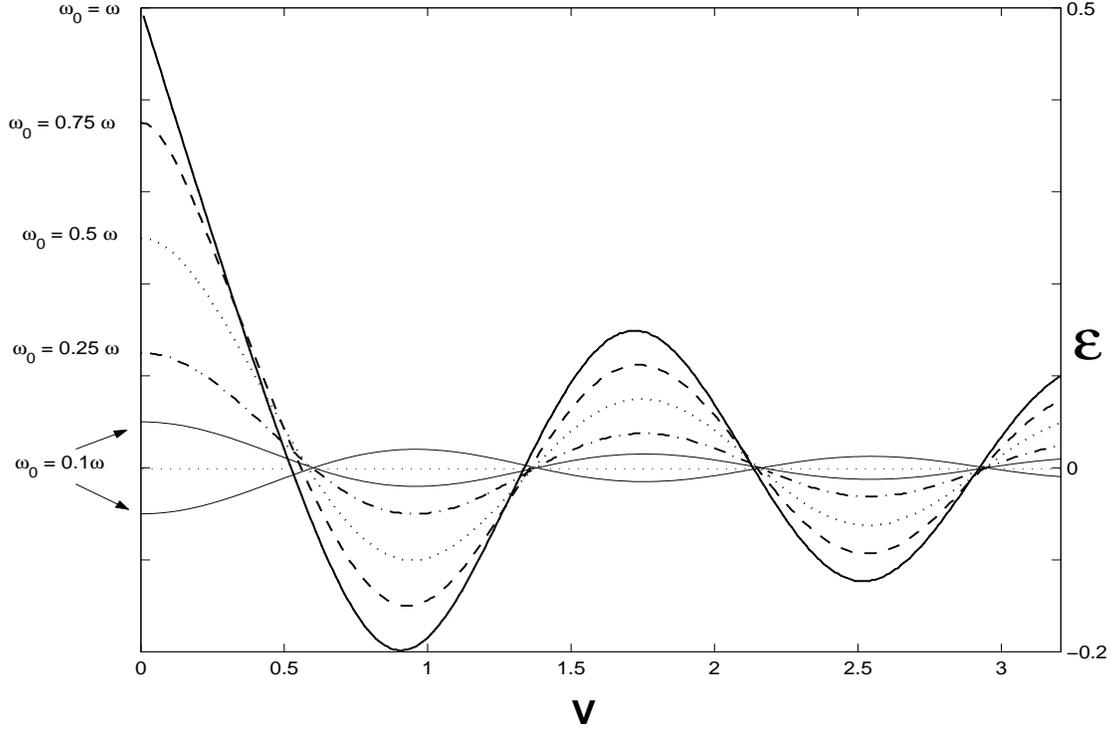,height=10cm,width=15cm}
\caption{Quasi-energy as a function of driving amplitude ($v$), for different values of $\wo \equiv \omega_0 /\omega \leq 1$ (high frequencies) . }
\end{center}
\end{figure}

\begin{figure}
\begin{center}
\epsfig{figure=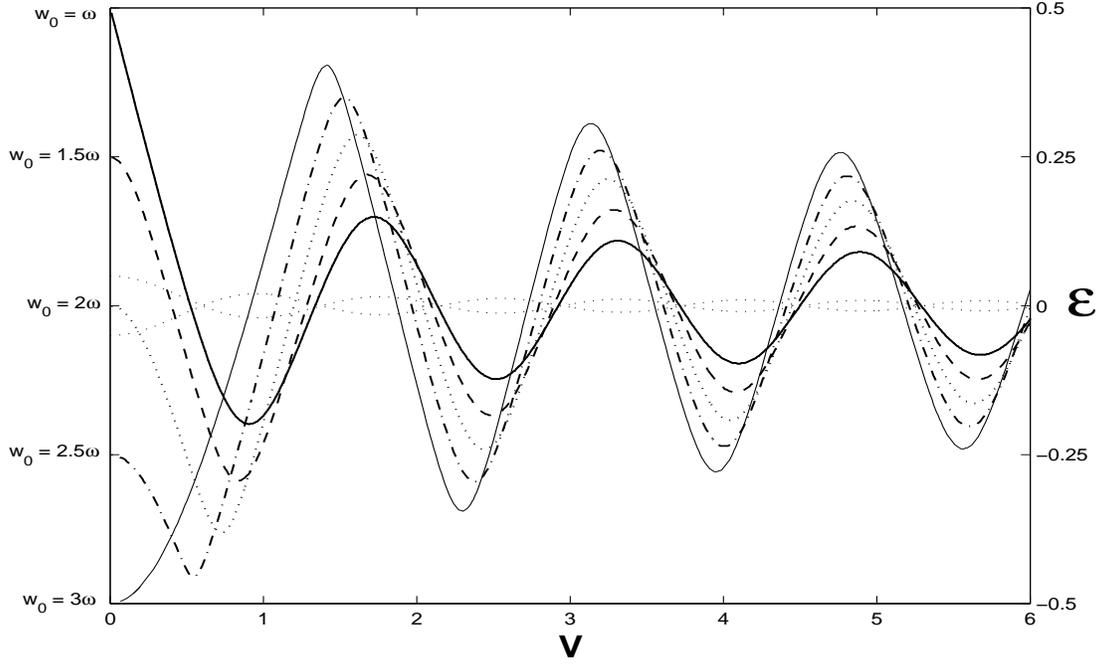,height=9cm,width=15cm}
\caption{Quasi-energy as a function of driving amplitude, for different values
of $\wo \geq 1$ (low frequencies). The functions $\pm 0.05 J_0 (4v)$ 
are plotted with dotted lines for comparison. }
\end{center}
\end{figure}

\newpage

\begin{figure}
\begin{center}
\epsfig{figure=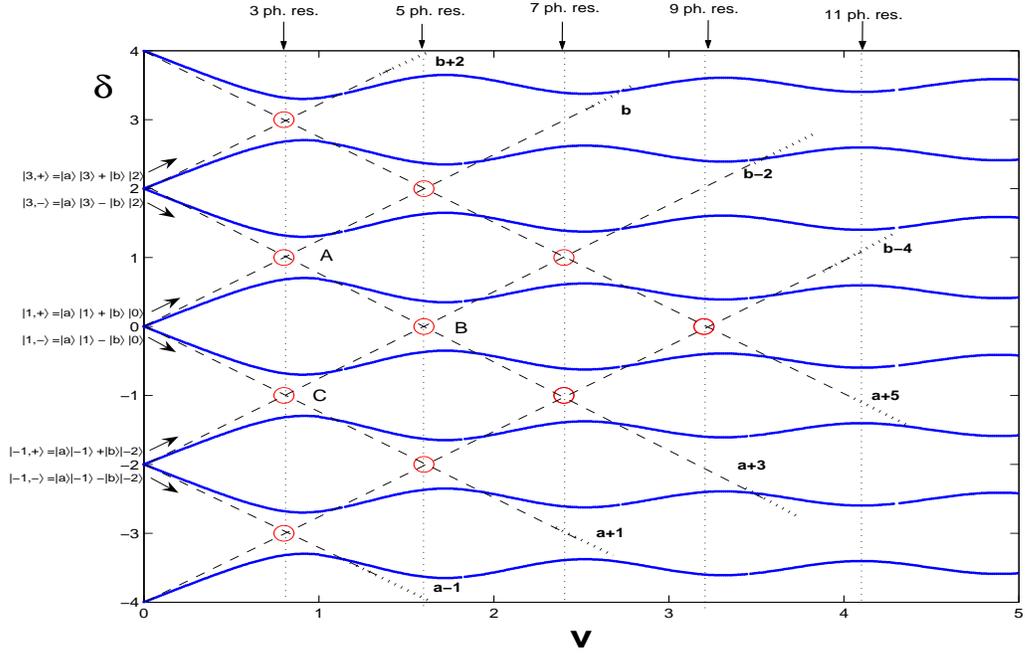,height=15cm,width=9cm,angle=-90}
\caption{Solutions of Eq. (20) as a function of driving amplitude for the
resonant case ($\wo = 1$). The avoided crossings structure can be accounted
for, in an approximate way, by a set of crossings of the eigenvalues
corresponding to the first-order FES indicated on the left of the
figure. Avoided crossings A and C are mainly due to a 3 photon
resonance between the corresponding FES; B corresponds to a 5 photon process.}
\end{center}
\end{figure}

\begin{figure}
\begin{center}
\epsfig{figure=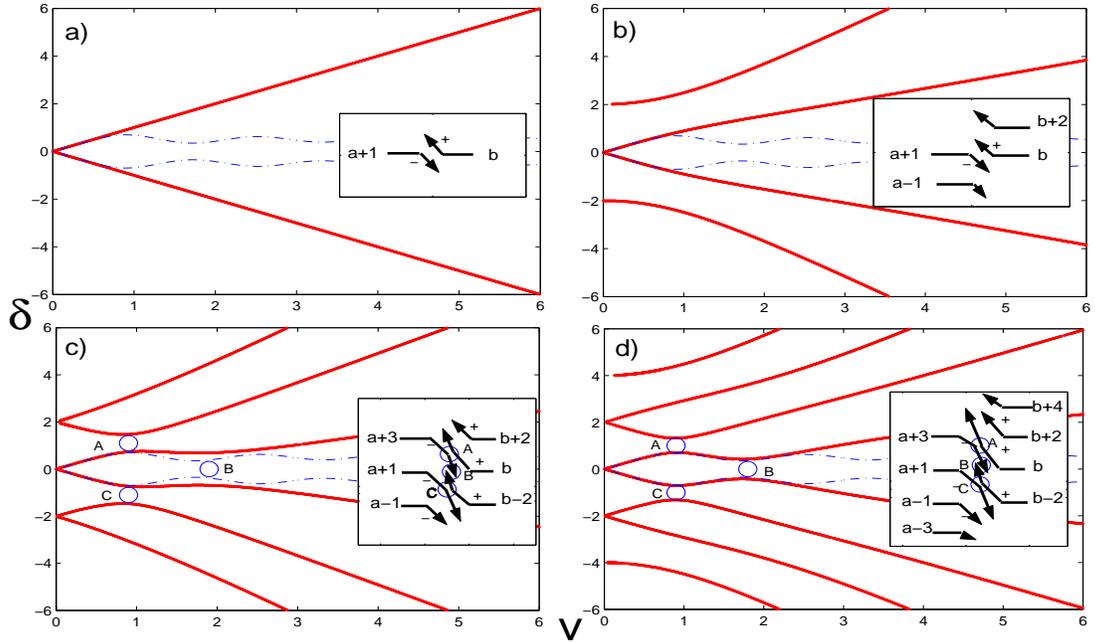,height=15cm,width=9cm,angle=-90}
\caption{Solutions of truncated versions of Eq. (22): $F_n^+ F_n^- =v^2$, for
different values of $n$. In a)
n=0; b) n=1; c) n=2; d)n=3. The labels $b-2n$ and
$a-2m-1$ correspond to terms in the CFs in Eq. (23) of the form $\delta+2n$ and
$\delta +2m+1+\wo$. As more terms are included new avoided crossings
appear. The $\pm$ signs are used to indicate each one of the two solutions
that originate at the points $\delta=2n$ in the y-axis.}
\end{center} 
\end{figure}

\begin{figure}
\begin{center}
\epsfig{figure=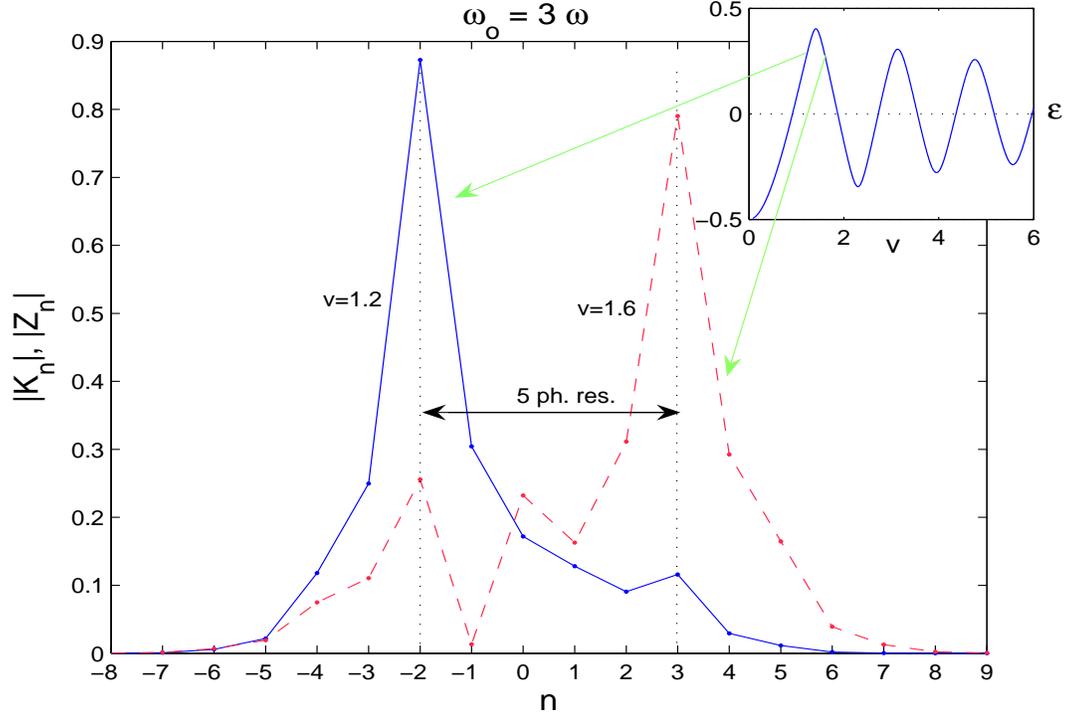,height=10cm,width=15cm,angle=0}
\caption{Components of the FES for $\wo=3$ and for two diffrent values of the
driving amplitude ($v=1.2,~1.6$). The avoided crossing that occurs near
$v \sim 1.4$ is due to a 5 photon process between two first-order FES.}
\end{center}
\end{figure}

\begin{figure}
\begin{center}
\epsfig{figure=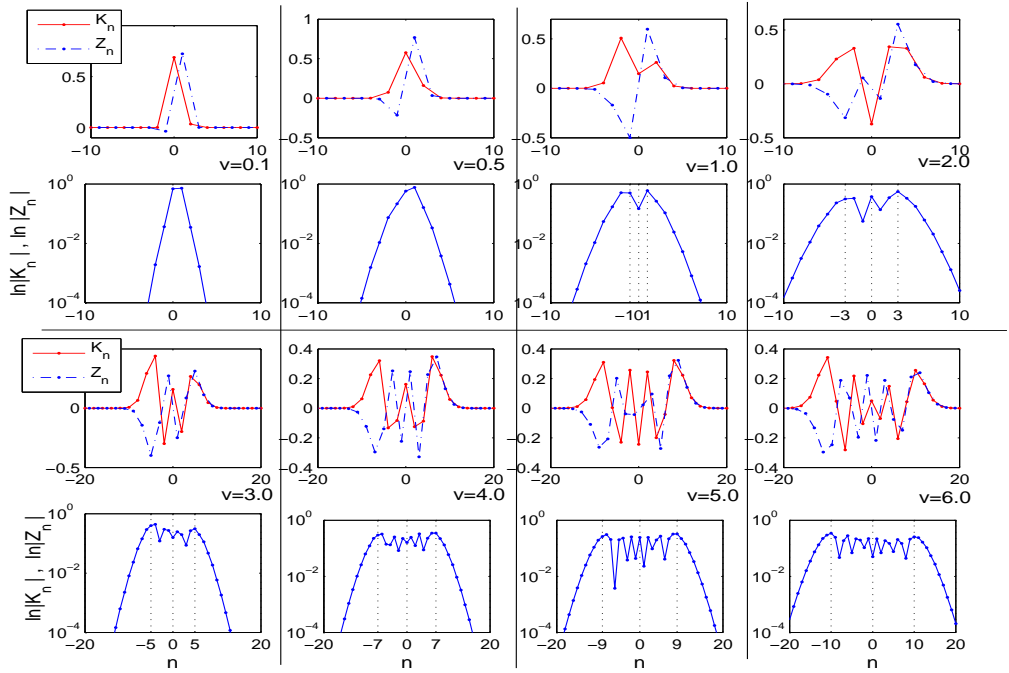,height=15cm,width=10cm,angle=-90}
\caption{Components of the FES for $\wo=1$ and for eight different values of the
driving amplitude. The logaritmic plots show the "plateau" structure of
the FES, which gives rise to the well-known plateau studied in HHG. The
cut-off location ($N_e$) follows the approximate rule $N_e = 2v-1$.}
\end{center}
\end{figure}

\begin{figure}
\begin{center}
\epsfig{figure=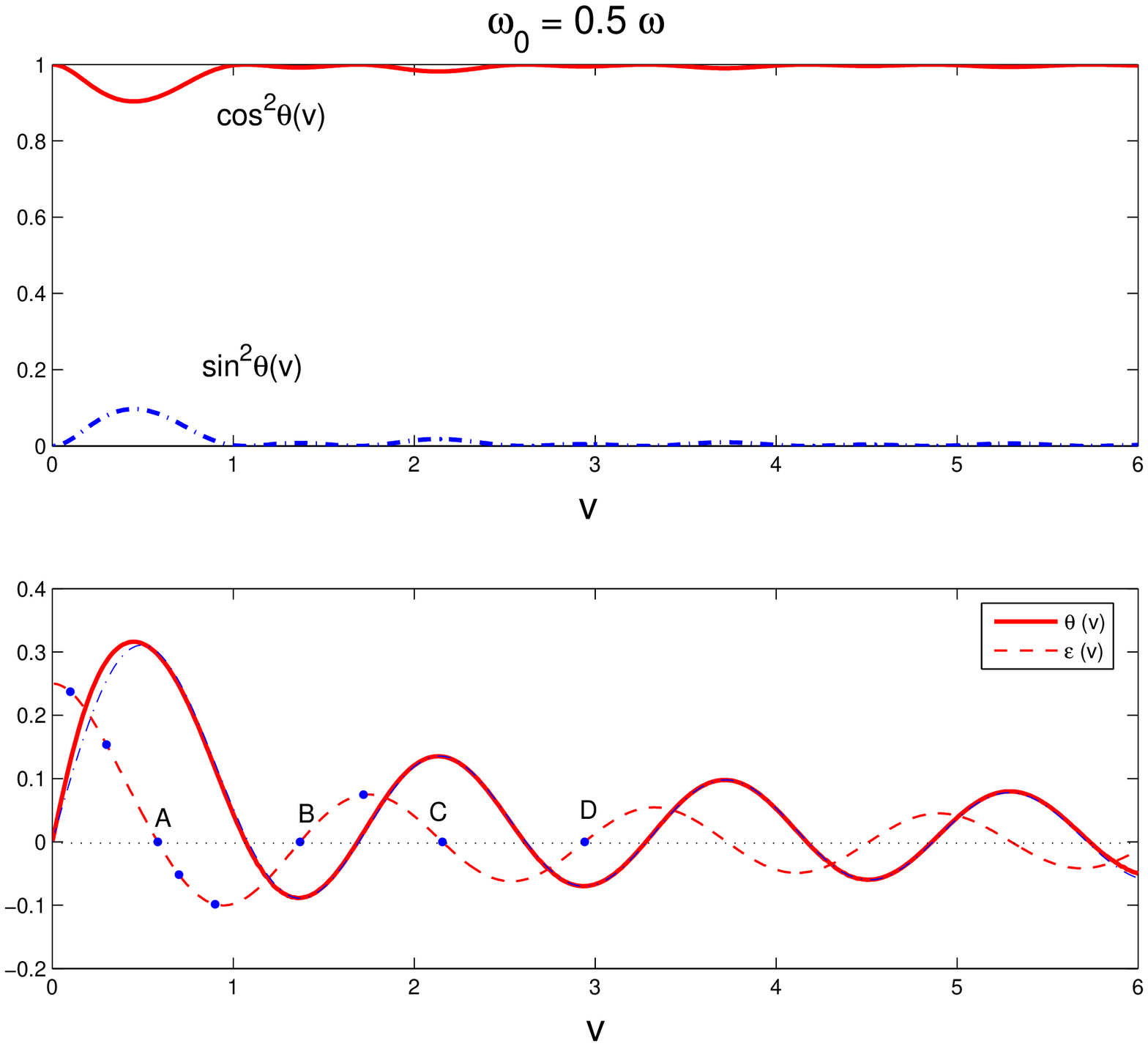,height=10cm,width=15cm,angle=0}
\caption{The upper panel shows the functions $\textrm{cos}^2 (\theta (v))$ and
$\textrm{sin}^2 (\theta (v))$, for $\wo=0.5$. 
 At $t=0$, the FES is given by $|\phi^{-\eps} (0)\bra =
\textrm{cos}\theta (v)\ka + \textrm{sin}\theta (v)\kb$. The lower panel shows
the functions $\theta (v)$ (solid line)  and $\eps (v)$ (dash line). The
dash-dot line corresponds to the Struve function $\wo\frac{\pi}{4}H_0 (4v)$.
Points A, B, C, D, correspond to the values of $v$ for which DL occurs.}
\end{center}
\end{figure}

\begin{figure}
\begin{center}
\epsfig{figure=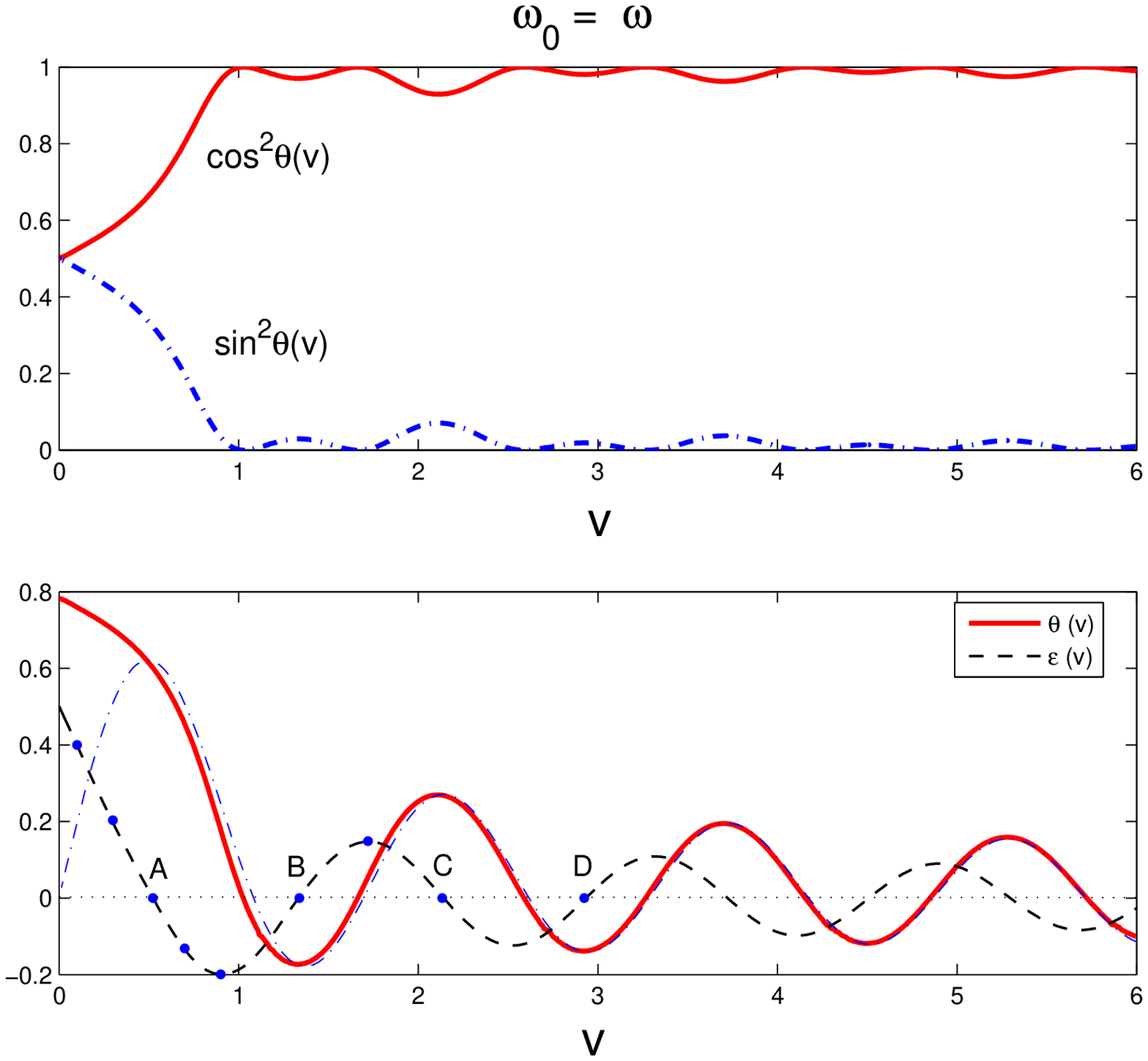,height=10cm,width=15cm,angle=0}
\caption{Same as in Fig. 7. Here $\wo=1$.}
\end{center}
\end{figure}

\begin{figure}
\begin{center}
\epsfig{figure=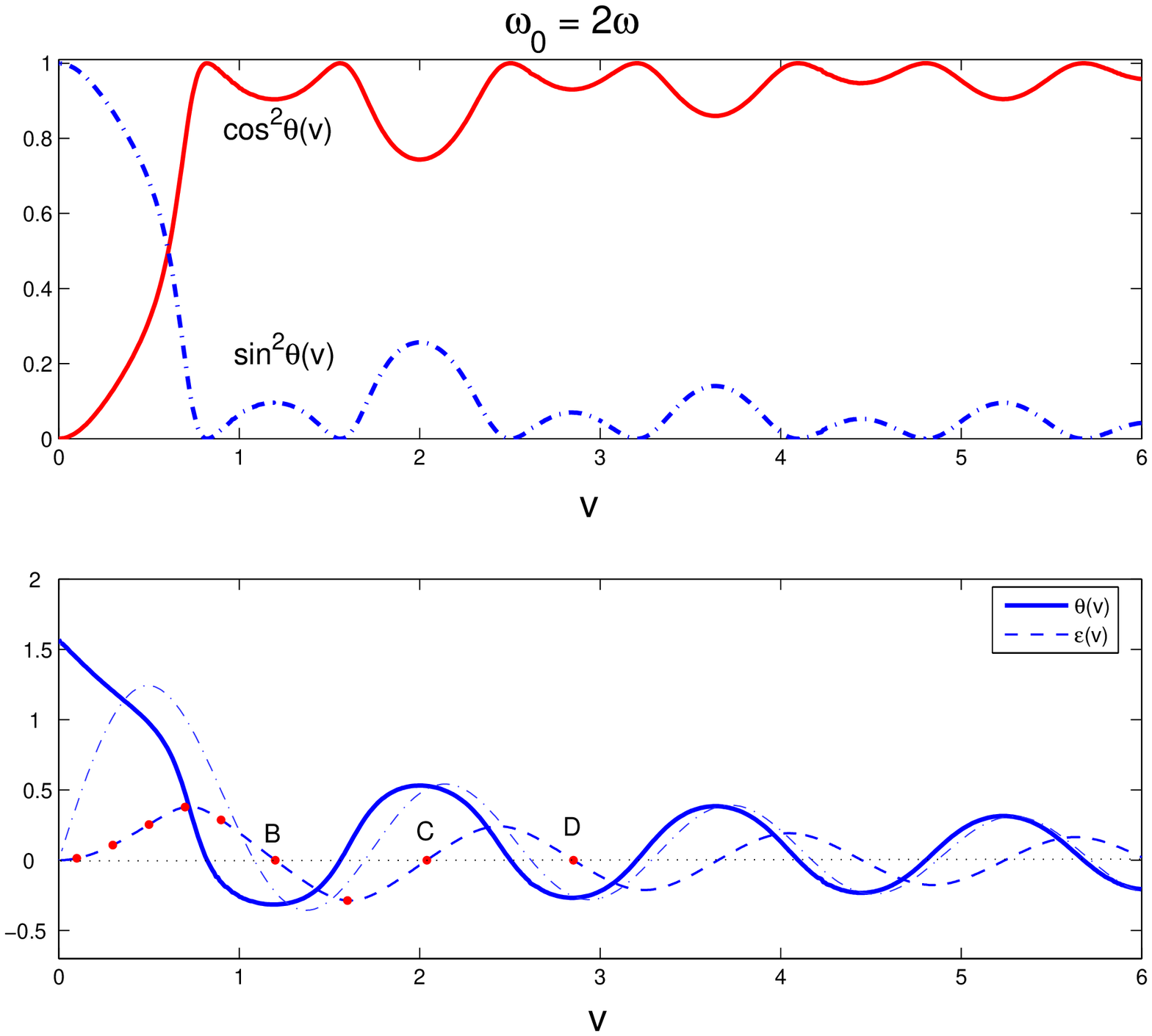,height=10cm,width=15cm,angle=0}
\caption{Same as in Figs. 7,8. Here $\wo = 2$.} 
\end{center}
\end{figure}

\begin{figure}
\begin{center}
\epsfig{figure=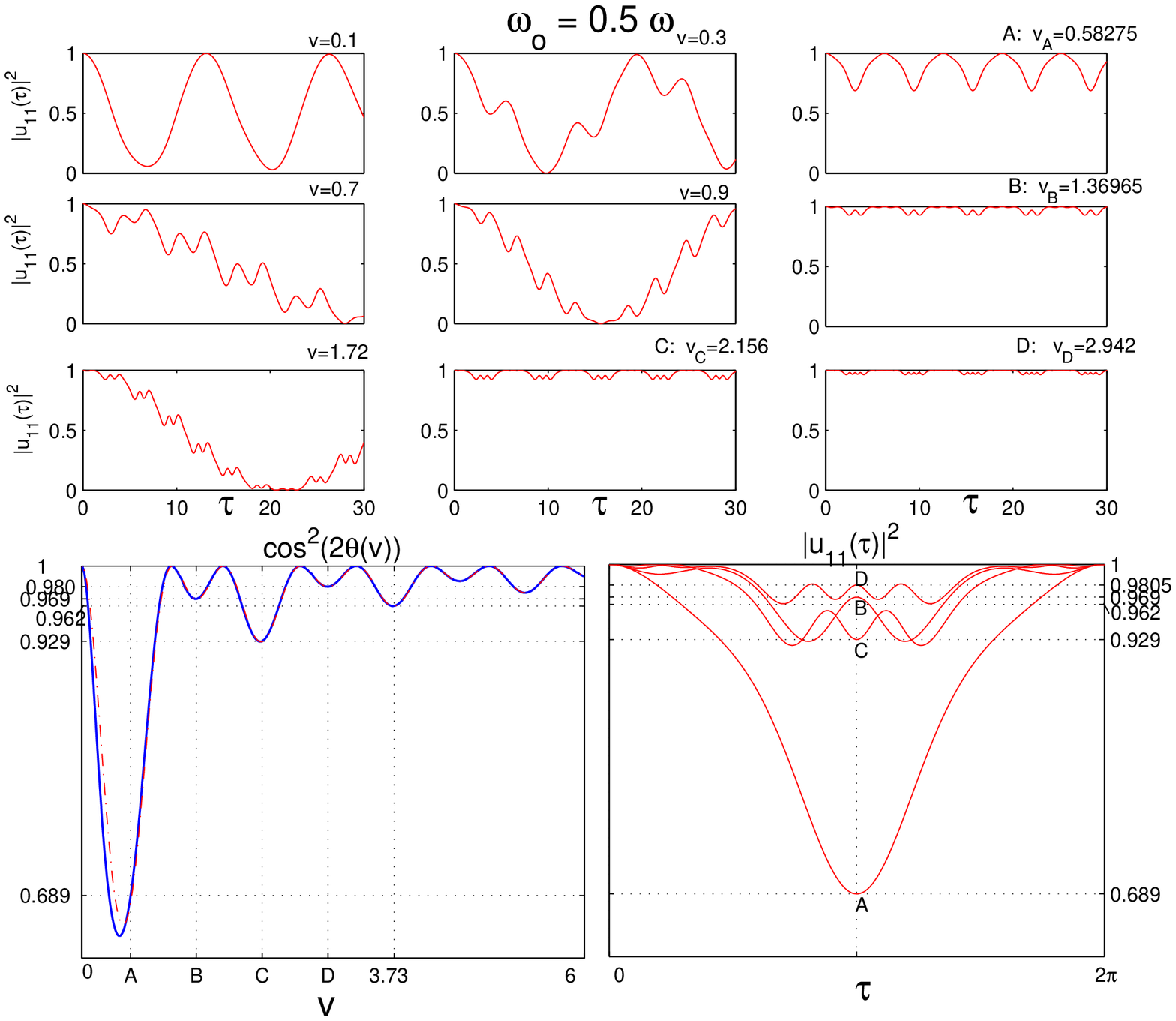,height=15cm,width=10cm,angle=-90}
\caption{The probability to remain in the initial state $|1\bra$ as a function of
time ($\tau = \omega t$), for 9 different values of the driving
amplitude and for $\wo=0.5$. The lower panels show that the function $\cttv$,
 at the DL points (A,B,C,D), gives a good approximation to the amount of
localization $P_l$ in this system. $P_l = \textrm{min}(|u_{11} (\tau)|^2)$. In
the lower left panel the function $\textrm{cos}^2 (2\wo\frac{\pi}{4}H_0 (4v))$
is shown with a dash-dot line for comparison.}
\end{center}
\end{figure}

\begin{figure}
\begin{center}
\epsfig{figure=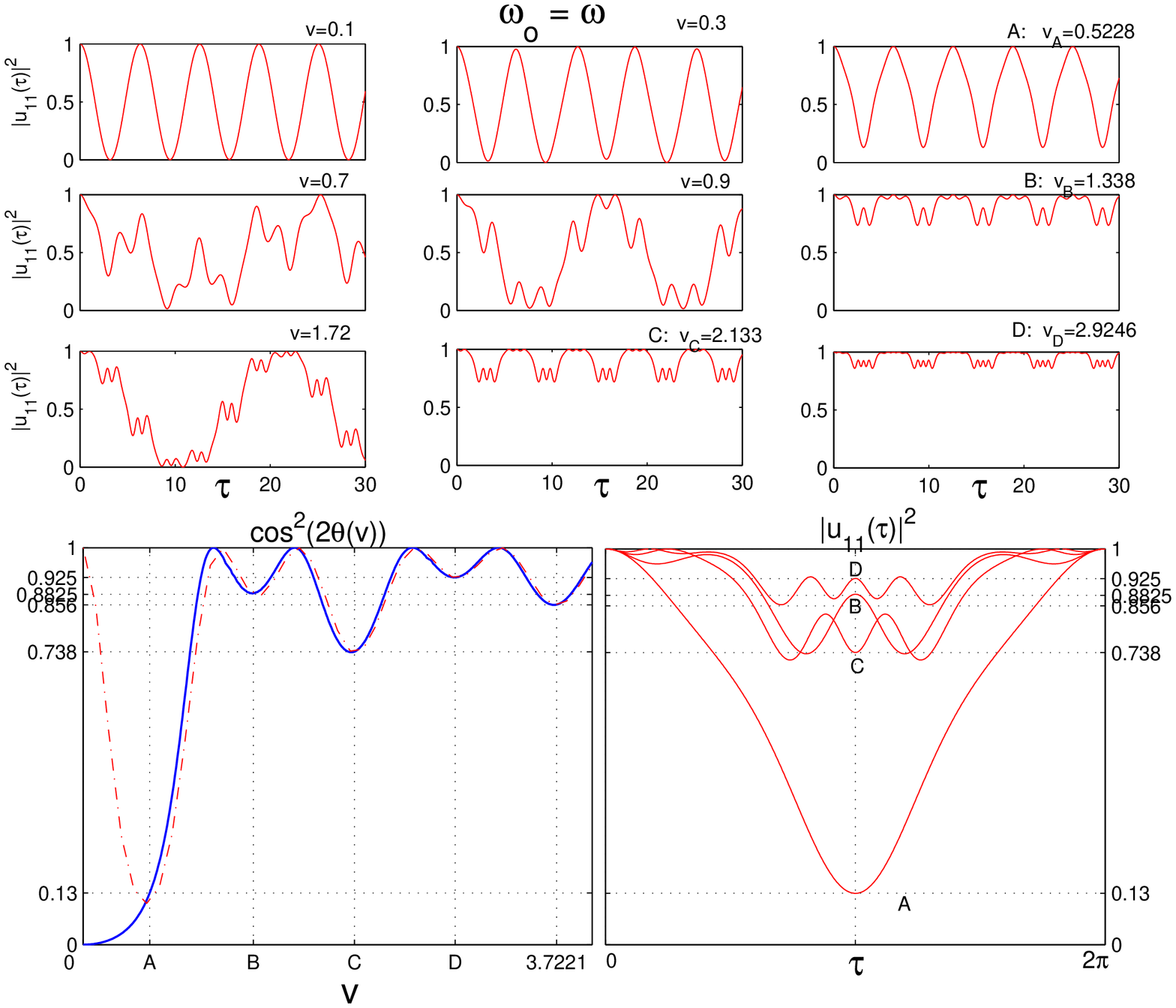,height=15cm,width=10cm,angle=-90}
\caption{FIG.11. As in Fig.10 . Here   $\wo =1$.}
\end{center}
\end{figure}

\begin{figure}
\begin{center}
\epsfig{figure=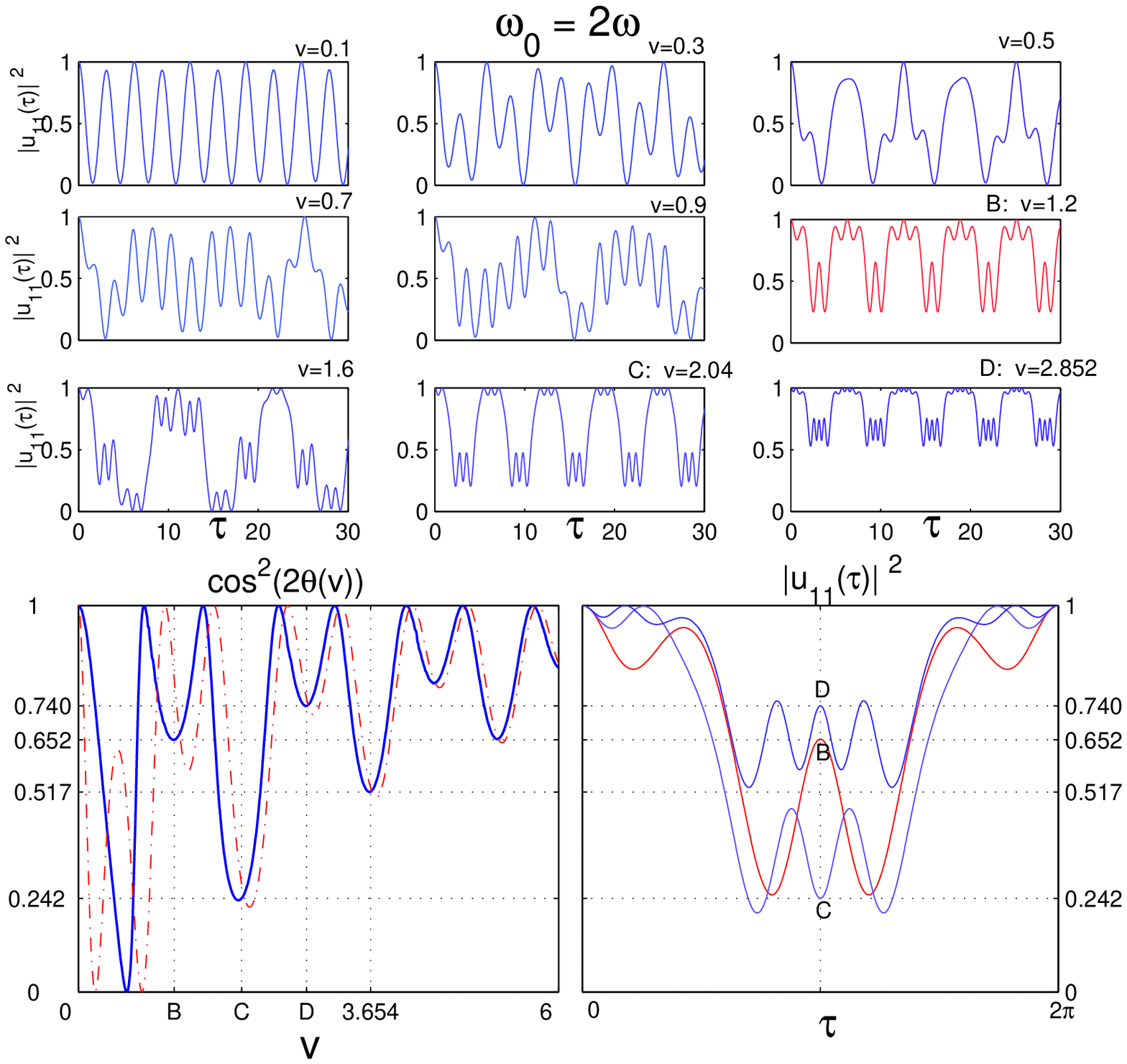,height=15cm,width=10cm,angle=-90}
\caption{As in Figs.10,11. Here  $\wo=2 $.}
\end{center}
\end{figure}

\begin{figure}
\begin{center}
\epsfig{figure=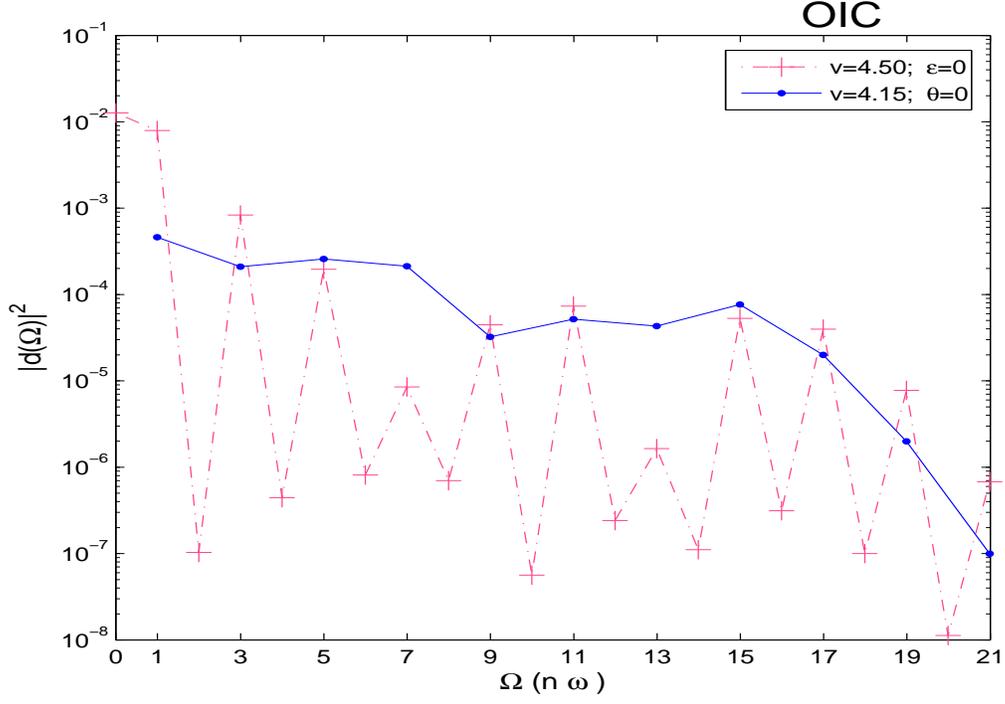,height=10cm,width=15cm,angle=0}
\caption{The emission spectrum of the system, for the "Optical" Initial
Condition (OIC) and for two values of the driving amplitude:  $v=4.5$
corresponds to a DL point and $v=4.15$ gives $\theta(4.15) =0$. The Fourier components of $\ket
d(t)\bra$ are connected with a continuous line for better visualization.}
\end{center}
\end{figure}

\begin{figure}
\begin{center}
\epsfig{figure=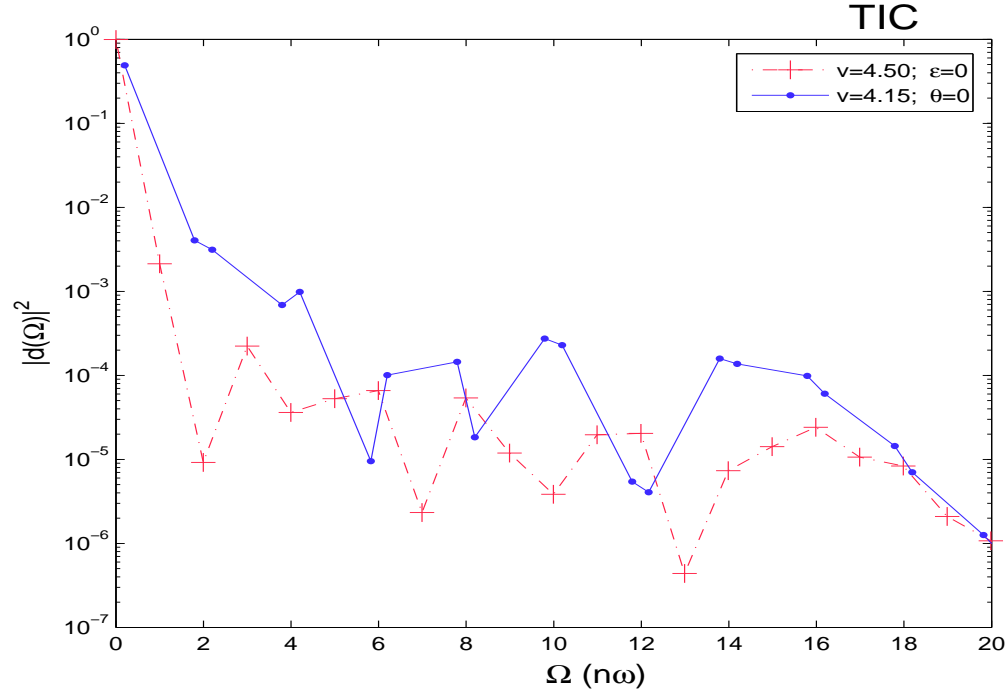,height=10cm,width=15cm,angle=0}
\caption{Same as in Fig. 13. Here for "Tunneling" initial condition
  (TIC).}
\end{center}
\end{figure}

\begin{figure}
\begin{center}
\epsfig{figure=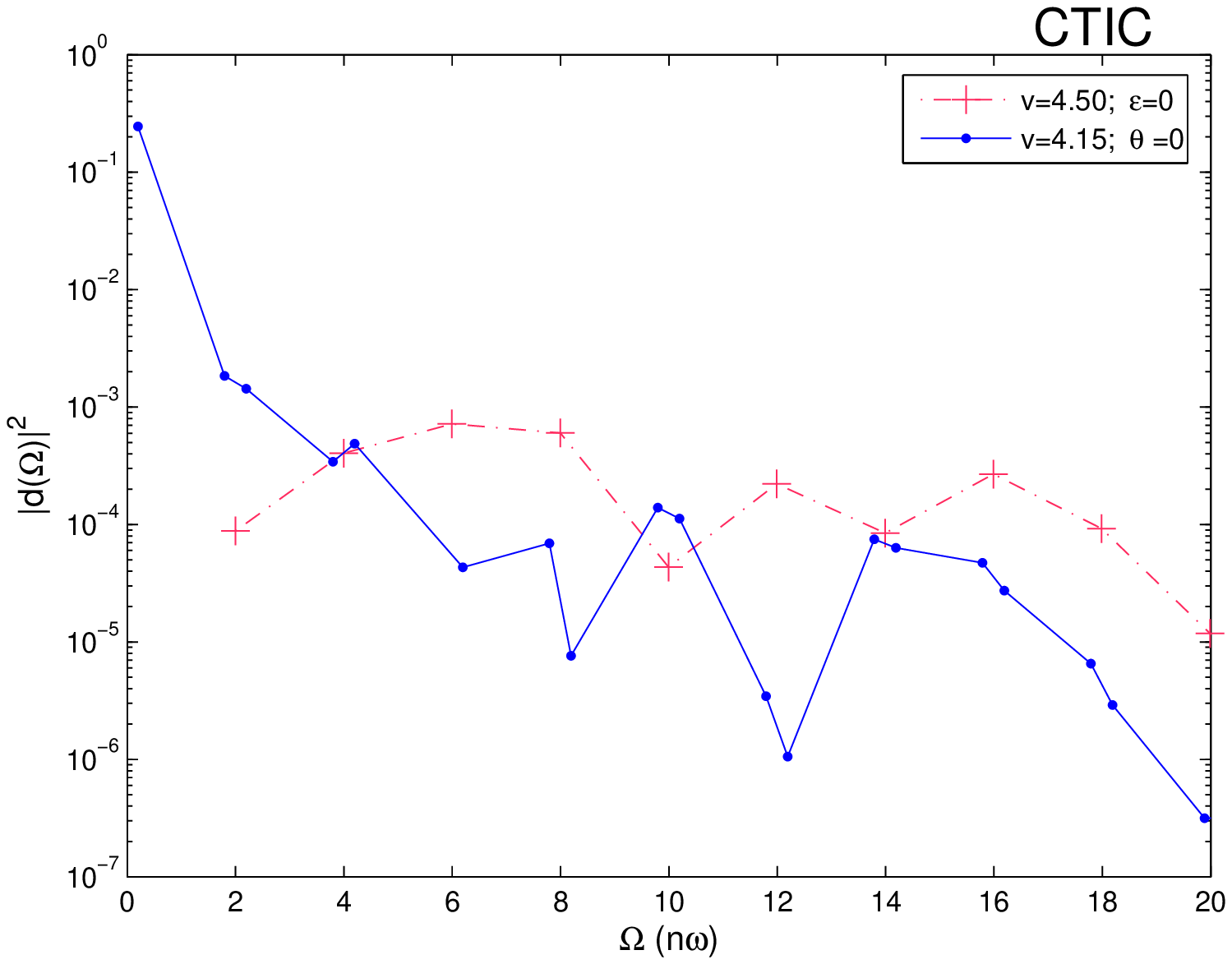,height=10cm,width=15cm,angle=0}
\caption{Same as in Fig. 13,14. Here for  "Complex Tunneling" Initial
Condition (CTIC).}
\end{center}
\end{figure}


\begin{thebibliography}{aa}
\bibitem{approx}For the resonant case, we have obtained a
numerical fit, that can be used for any value of $v$. It was obtained by
displacing the first couple of zeros of the Bessel
function $J_0(4v)$ to the left, and linearizing the initial behavior by
multiplying this function by a suitable factor. The resulting
expression that approximates the behavior of
$\epsilon (v)$ to two decimal points is
$\epsilon (v) \simeq \frac{1}{2}(0.5353+1.7(v-0.5228)^2)\frac{
1+(2v)^4}{1+(0.5353+1.7(v-0.5228)^2)(2v)^4}J_0
(4v+\frac{0.6}{1+5(v-0.5228)^2}v)$
\bibitem{convergence} Clearly, in the limit $n\rightarrow \infty$, $F^\pm _n
(E)\rightarrow F^\pm  (E)$. Even though the convergence of
continued fractions is a complex mathematical subject, we have not
yet found a continued fraction related to a physical system which
failed to have a limit. For all calculations that involve a C.F.
in this paper, convergence to any specified precision was always
achieved.
\end{thebibliography}
\end{document}